\documentclass[12pt]{article}
\usepackage[hmargin=2.9cm,vmargin=3.2cm]{geometry}
 \usepackage{graphicx}
 \usepackage{amsmath}
 \usepackage{amssymb}
  \usepackage{enumerate}
\usepackage{cite}
\usepackage{mathabx}
\usepackage{url}
\newcommand{\bey}{\begin{eqnarray}}
\newcommand{\eey}{\end{eqnarray}}
\usepackage{etoolbox}

\usepackage{lipsum}

\let\OLDthebibliography\thebibliography
\renewcommand\thebibliography[1]{
  \OLDthebibliography{#1}
  \setlength{\parskip}{0pt}
  \setlength{\itemsep}{2pt plus 0.3ex}
}

\newcommand{\R}{\mathbb{R}}

\begin{document}
\title{Quantum Field Theory based Quantum Information: Measurements and Correlations}
\author {Charis Anastopoulos $^{(a)}$ \thanks{anastop@upatras.gr} ,
Bei-Lok Hu $^{(b)}$ \thanks{blhu@umd.edu} ,
and   Konstantina Savvidou $^{(a)}$ \thanks{ksavvidou@upatras.gr}
 \\ \\
\noindent $^{(a)}$  {\small Department of Physics, University of Patras, 26500 Greece} \\
\noindent $^{(b)}$ {\small Maryland Center for Fundamental Physics and Joint Quantum Institute},\\ {\small  University of Maryland, College Park, Maryland 20742-4111 U.S.A.}}

\maketitle

\begin{abstract}
 This is the first in a series of papers aiming to develop a relativistic quantum information theory in terms of unequal-time correlation functions in quantum field theory.
In this work, we highlight two formalisms which together can provide a useful theoretical platform suitable for further developments: 1) Quantum field measurements using the Quantum Temporal Probabilities (QTP) method; 2) Closed-Time-Path (CTP) formalism for causal time evolutions.
QTP incorporates the detector into the quantum description, while emphasising that the records of measurement are macroscopic and can be expressed in terms of classical spacetime coordinates. We first present a new, elementary derivation of the QTP formulas for the probabilities of n measurement events. We then demonstrate the relation of QTP with the Closed-Time-Path  formalism, by writing an explicit formula that relates the associated generating functionals. We exploit the path integral representation of the CTP formalism, in order to express the measured probabilities in terms of path integrals. After this, we provide some simple applications of the QTP formalism. In particular, we show how Unruh-DeWitt detector models and Glauber’s photodetection theory appear as limiting cases. Finally, with quantum correlation being the pivotal notion in relativistic quantum information  and measurements, we highlight the role played by the CTP two-particle irreducible effective action which enables one to tap into the resources of non-equilibrium  quantum field theory for our stated purpose.


\end{abstract}

\section{Introduction}

Quantum information theory (QIT) is a quantum extension of classical information theory.  The primary force behind the explosive developments of QIT during the last decades is entanglement, the uniquely quantum feature which provides the new and powerful resource for quantum computing,  quantum communication, quantum metrology and more.

While potential applications of quantum information   technologies have drawn close attention of government and private research and development, the theoretical foundation of QIT is surprisingly lacking in a major portion. Our understanding of QIT is far lagging behind the fully developed quantum theory of nature, namely, quantum field theory (QFT), which has proven its validity and worth in the full range of physical sciences from particle-nuclear physics to atomic, optical and condensed matter physics, from quarks to black holes and the early universe.  So far, quantum information theory has been largely developed in the context of non-relativistic quantum mechanics, which is a small corner of full QFT.  It is ostensibly inadequate when basic relativistic effects like locality, causality and spacetime covariance, need be accounted for.

\subsection{QITs not based on QFT are incomplete}

QFT is a quantum theory that also includes axioms about the effect of spacetime structure on the properties of quantum systems, especially regarding the causal propagation of signals. In contrast, current quantum information theories do not incorporate the latter axioms. Their notion of causality, based on the sequence of successive operations on a quantum system, lacks a direct spacetime representation. As a result, current QITs   cannot make crucial relativistic distinctions, for example, between timelike and spacelike correlations, it does not describe real-time signal propagation, and it ignores relativistic constraints on permissible measurements. A genuinely relativistic QIT must overcome such limitations \cite{AnSav22}.

Furthermore, experiments that study causal information transfer or gravitational interaction in multi-partite quantum systems require a QFT treatment of interactions for consistency.  A non-QFT description may severely misrepresent the theoretical modeling of the system or the physical interpretation of the results. This point is crucial for tests of foundational issues of quantum theory invoking quantum information concepts such as entanglement and decoherence or methods such as signaling and communication. This is especially so for quantum  information  experiments in space \cite{Rideout, DSQL2} and for experiments designed to explore quantum effects from gravity \cite{AnHu15, Bose17, Vedral17, AnHu20}.

The introduction of the key concepts of spacetime covariance and causality in  QIT forces us to address problems that originate from the foundations of QFT. We draw an incomplete list of noteworthy issues below.

\medskip

\noindent {\em Quantum States.} In set-ups that involve more than two quantum measurements, the standard  state update rule implies that the
quantum state is genuinely different when recorded from different Lorentz frames  \cite{Bloch, Aharonov, PeTe}. There is no problem with the theory’s physical predictions that are expressed in terms of (multi-time) probabilities\cite{WH65}. However, the usual notions of quantum information (entropy, entanglement) are defined through the quantum state, and as such, they are ambiguous in relativistic measurement set-ups.

\medskip

\noindent {\em Local Operations.}   It is a challenging problem to formalize the notion of a localized quantum system in QFT. There are powerful theorems  demonstrating that even  unsharp localization in a spatial region leads to faster than light signals \cite{Malament,  Heg1}.  Hence, expressing the crucial quantum informational notion of a local operation in terms of spatial localization can lead to conflicts with relativistic causality.

\medskip

\noindent {\em Projective Measurements.}  There are strong arguments that ideal (i.e., projective) measurements in QFT are incompatible with causality \cite{Sorkin, BJK}, essentially because they change the quantum state over a full Cauchy surface. However, existing quantum information notions (even the very notion of a qubit) presuppose maximal extraction of information through ideal measurements.

\medskip

We contend that the development of consistent relativistic QIT requires a measurement theory that (i) respects causality and locality, and (ii) it is expressed in terms of quantum fields\footnote{The explicit formulation in terms of QFT is the point of divergence of our approach from past works on formulating a relativistic QIT  \cite{Peres2, PoVai, InstMeas, GoPr}.}. Furthermore, this measurement theory ought to be practical, i.e., it must provide non-trivial predictions for experiments that are accessible now or in the near future. For other reasons why a QFT measurement theory is needed, see Ref. \cite{Grimmer}.

\subsection{Past work on QFT measurements}

To the best of our knowledge, the earliest discussion of measurements on quantum fields was by Landau and Peierls \cite{LP31}. They derived an inequality for the localization of particles. Bohr and Rosenfeld criticized some of their assumptions \cite{BoRo}, and proved that the measurement of field properties  requires a test particle of macroscopic scale: the particle's  charge $Q$ must be much larger than the electron charge e.

Arguably the first explicit model for QFT measurements was Glauber's photodetection theory \cite{Glauber1, Glauber2}, which was developed as a quantum generalization  of the classical theory of electromagnetic  coherence. Glauber's theory  defines unnormalized probabilities for photon detection in terms of the electric field operators ${\bf E}(x) $  and the field's quantum state $|\psi\rangle$. The joint probability density $P(x_1, x_2, \ldots, x_n)$ for the detection of a photon at each of the spacetime points $x_1, x_2, \ldots, x_n$ is given by
\bey
P(x_1, x_2, \ldots, x_n)=  \langle \psi| E^{(-)}(x_1)E^{(-)}(x_2)  \ldots E^{(-)}(x_n) E^{(+)}(x_n) \ldots E^{(+)}(x_2) E^{(+)}(x_1)|\psi\rangle, \label{jointdet1}
\eey
where $E^{(+)}$ is the positive-frequency component  and $E^{(-)}$ the negative-frequency component of the projected field vector field ${\bf n}\cdot{\bf E}(x)$.
The probability density (\ref{jointdet1})  is essential for the definition of higher-order coherences of the electromagnetic field, and consequently, for the description of phenomena like the Hanbury-Brown-Twiss effect,   photon bunching and anti-bunching \cite{QuOp}.

The success of Glauber's theory in quantum optics has been immense. However, its applicability is limited to measurements of photons. Another problem is that the field splitting into positive and negative frequency components is non-local. The probability density (\ref{jointdet1}) assumes the   Rotating Wave Approximation  for the interaction of the field to the detector \cite{RWA1, RWA2}. This approximation
misrepresents the retarded propagation of the electromagnetic field. This means that the model  may face problems with causality in set-ups that involve propagation along long distances \cite{phc1, phc2, FuMa}.

A very common class of models for QFT measurements are based on the notion of {\em Unruh-DeWitt}  detectors \cite{Unruh76, Dewitt}. These models first appeared in the study of the Unruh effect, in order to demonstrate the effects of acceleration on the quantum field vacuum. In an Unruh-DeWitt detector, the quantum field is coupled to a point-like   system that  moves along a pre-determined spacetime trajectory  $x(\tau)$, where $\tau$ is the trajectory proper-time. An Unruh-DeWitt detector model  essentially describes a 0+1 field theory interacting with a 3+1 field theory  through an interaction that is defined  by  and embedding of the 0+1 spacetime to the 3+1 spacetime (i.e., the detector's trajectory). Hence, the detector is kinematically pointlike: physical observables by necessity are defined along the trajectory. However, the dynamics may
 well  involve field couplings that correspond to an extended system \cite{LoSa, Perche}.

  While a first-principles derivation of the detector's response for non-inertial  motion  is still missing, Unruh-DeWitt detector models are the simplest models for describing QFT measurements \cite{GGM22}, and they have found several applications---for a  sampling of the latter, see Ref. \cite{HLL12}. Unruh-DeWitt detectors are limited in that  detector degrees of freedom are not described by  a QFT.
Non-pointlike Unruh-DeWitt detectors may lead to non-causal signals in set-ups that involve multiple detectors. For different perspectives about causality in Unruh-DeWitt detectors, see Refs.
 \cite{sl1, sl2,   sl4}.

 Measurement models have also been constructed in the context of  algebraic QFT \cite{HeKr, OkOz, Dop, FeVe, Few19, Ruep}, primarily in order to address issues of causality and locality. For example, Ref. \cite{FeVe} considers a system and a probe / apparatus, both described by a QFT. The two field systems start separated and   interact within a bounded spacetime region. In Minkowski spacetime, this interaction is described by an S matrix, and it leads to correlations between observables on the system and records on the apparatus.
  Then, one defines probabilities for the latter in terms of operators that are well defined on the probe's Hilbert space. The overall analysis is fully consistent with QFT. Furthermore, it also works for curved spacetimes and it is not tied to the existence of an S-matrix, i.e., the existence of asymptotic in-out regions. It is straightforwardly generalized to sequences of  measurements \cite{FJR22}.    However, this approach  has not  yet been developed into a practical tool capable of concrete physical predictions.

\subsection{Our approach}

In this work, we describe QFT measurements using the Quantum Temporal Probabilities method \cite{QTP1, QTP2, QTP3}. The original motivation of this method was to provide a general framework for  temporally extended quantum observables \cite{AnSav06, AnSav08, An08},  hence, the name.
The key idea  in QTP is to distinguish between the time parameter of Schr\"odinger's equation from the time variable associated to  particle detection \cite{Sav99, Sav10}. The latter is then treated as  a macroscopic quasi-classical variable associated to the detector degrees of freedom. Here, we use the word `quasi-classical' as in the decoherent histories approach to quantum theory \cite{Gri, Omn1, Omn2, GeHa1, GeHa2, hartlelo}. A quasi-classical variable is a coarse-grained quantum variable that  satisfies appropriate decoherence conditions, and as a consequence its dynamics can be approximated by  classical evolution equations  \cite{GeHa2, hartlelo}. Hence, the detector admits a dual description: in microscopic scales it is described   by quantum theory, but its macroscopic records are expressed  in terms of classical spacetime coordinates. Furthermore, in QTP the detector is also described in terms of quantum fields. Its interactions with the detector respect causality, because the interaction Hamiltonian  is a local functional of  quantum fields.

QTP provides room for sufficient freedom when dealing with the model for the apparatus, when compared with other approaches. However, at the current practical applications of the framework, it converges with the results of other approaches. The current emphasis is on the derivation of probabilities for measurements, while incorporating the effects of the apparatus in a small number of phenomenological quantities that can be related to experiments.

 Glauber's detection theory and Unruh-DeWitt detector models emerge from QTP as limiting cases that characterize specific regimes: Glauber's theory appears at the limit of very small characteristic timescales for the detector, while Unruh-DeWitt models appear at the limit of very short length-scales. In comparison to the algebraic QFT approaches to measurements, QTP provides the same results to leading order in perturbation theory, but allows for the definition of observables for the spacetime observables, and it is embedded within a nuanced analysis of the quantum-classical transition in the detector through the decoherent histories approach.

A key insight from QTP is that probabilities for measurements are defined in terms on specific unequal-time
correlation functions of the quantum field. This is particularly important, as such correlation functions are a staple of QFT.  Powerful  methods  have been developed for their calculation and the analysis of their properties. The specific correlation functions relevant to QTP are not the usual ones of S-matrix theory (in-out formalism), but they appear in the Closed-Time-Path (CTP) (Schwinger-Keldysh or `in-in') formalism \cite{ctp1, ctp2, ctp3, ctp4, ctp5}. The CTP formalism improves over the in-out formalism, in that it allows for causal equations of motion, and it has found many applications in early universe cosmology \cite{ch87, wein05}, nuclear-particle process \cite{ctp5, CH08, Berges, Berges2} and condensed matter physics \cite{coma1, coma2}.

The connection that we establish between the two formalisms allows us to translate between the concepts  of quantum measurement theory and of quantum field theory. The former is based on operational notions like Positive-Operator-Valued measures (POVMs) and effects, while the latter employs functional methods like path integrals in order to express its predictions in a manifestly covariant language. In this work, we show, for example, how to express POVMs for particle detection in terms of path-integrals.

A key point of our analysis is the central role played by the unequal-time correlation functions. In the QTP description, they contain all information about measured probabilities. The detection probability for $N$ events depends on an $2N$-unequal time correlation function. This has the following implication. Two-event measurements typically characterize bipartite systems, and as such the corresponding probabilities are related to entanglement. Hence, field four-point functions contain all information that pertains to bipartite entanglement. A relation at the level of non-relativistic fields was shown   in \cite{QTP3}.

An analysis at the level of the correlation functions brings us closer to the main ideas of non-equilibrium QFT, where irreversibility is implemented by the truncation of the hierarchy of field  correlation functions, and the slaving of higher correlation functions to the two-point function. Again, there is a natural relation between QTP and non-equilibrium formalisms that are based on CTP.  We argue that the QTP probabilities function as a registrar of information for the quantum field, as it keeps track  of how much information resides in which level of correlation functions, and how it flows from one level to the other during dynamical evolution.

We believe that the scheme presented here has good potential to systemize quantum information in QFT,   and to identify the parts of this information that is relevant to the field's statistical, stochastic and thermodynamic behavior.
  Hence, this formalism could provide a concrete method for defining  quantum information in QFT via the correlation hierarchy, as has been proposed in \cite{Erice95}. Such a definition would be very different from definitions of information in standard QIT that is based on the properties of the single-time quantum state.

\subsection{Our results}

Our results are the following.

First, we present a new derivation of the QTP  formulas  for the probabilities of $n$ measurement events. The derivation is intuitive and pedagogical, as it only requires elementary perturbation theory. It reproduces the results of the more rigorous QTP analysis to leading order  with respect to the system-apparatus coupling. This suffices for most applications. Then we  analyze the mathematical structure of the probability formulas. In particular, we show how they provide an explicit connection between the concepts of quantum measurement theory and QFT concepts, namely, unequal-time field correlation functions.

Second, we demonstrate the  relation of QTP with the Closed-Time-Path formalism, by deriving an explicit formula that relates the associated generating functionals.  The Closed-Time-Path formalism has a natural path integral formulation, which allows us to express the QTP probabilities in terms of path integrals. Conversely, QTP provides a theory of observables appropriate to the in-in formalism.

Third, we provide some demonstrative applications of the formalism. We briefly review the construction of time-of-arrival probabilities for relativistic particles. Then we show how the commonly employed models of Glauber's photodetection and Unruh-DeWitt detectors appear  as limiting cases of the QTP formalism. This implies, in particular, that there exists regimes in which QTP provides different predictions from Glauber's theory, and these differences may be experimentally distinguishable.

Fourth, we explore the links between QTP and non-equilibrium QFT. For field states with a large number of particles, QTP probabilities provide natural definitions for  Boltzmann-type observables. They also allow us to explore how information flows between the different levels of the Schwinger-Dyson hierarchy of correlation functions.

This paper is structured as follows. In Sec.  2, we present the main ideas of the QTP method, and we derive the probability formula for $n$-detection events. In Sec. 3, we derive the connection with the CTP formalism, and also derive a path integral expression for QTP probabilities. In Sec. 4, we present some applications of the formalism, and in Sec. 5, we explore the links to non-equilibrium QFT. In Sec. 6, we summarize and discuss our results. The Appendix A contains a conceptually more rigorous derivation of the QTP probability formula.

\section{Probabilities for QFT measurements }
In this section, we present the QTP approach to quantum field measurements. First, we explain in more detail the need for a QFT measurement theory. Then, we derive the QTP probability formulas and we analyze their structure. The key feature of the probability formula is that probability densities pertaining to $n$ measurement events are linear functionals of specific $2n$ unequal-time correlation functions.

\subsection{The need for a QFT measurement theory}

Most  QFT predictions involve set-ups with a single state preparation and a single-detection event, which can be described in terms of the S-matrix. For example,  S-matrix amplitudes determine scattering cross-sections;  S-matrix poles determine the spectrum of composite particles and decay rates. This gives the general impression  that there is no further need for an elaborate measurement theory.

However, this over-simplified view is deceptive. There are at least two cases where the S-matrix analysis is insufficient. First, in  quantum optics, we need joint detection probabilities  in order to describe phenomena like photon bunching and anti-bunching \cite{QuOp}. A first-principles calculation of joint probabilities requires a non-trivial description of quantum measurements. In non-relativistic physics, such a calculation involves the use of the state-update rule, i.e., quantum state reduction.
In QFT,  a universal rule for reduction is missing, it is absent even from axiomatizations\footnote{Rules for reduction can be obtained {\em a posteriori} if a QFT measurement scheme provides probability distributions for multiple measurements: see \cite{QTP2} for such definitions in QTP, \cite{GGM22} for the Unruh-DeWitt detectors and \cite{FeVe, Jubb} in the algebraic QFT approach.}.
 In practice, joint detection probabilities relevant to experiments are constructed through photodetection models, like Glauber's, whose derivation is rather heuristic.
  However, planned experiments in deep space \cite{Rideout,  DSQL2} that involve  measurement of electromagnetic field correlations at large separations  will arguably require a first-principles analysis of joint probabilities in order to take into account  the relative motion of detectors and delayed propagation.

The second case where S matrix theory does not suffice is when the expectation values of physical quantities at finite moments of time are called for, rather than just the asymptotic amplitudes offered in the `in-out' S-matrix theory.  A notable example is the construction of  dynamical equations for the evolution of spacetime metric in the early universe and in black holes from the solutions to the semiclassical Einstein equation \cite{nfsg, HM3}; these equations  require the evaluation of the expectation values of the stress-energy and its fluctuations.
Wherever causal time evolution is needed, rather than transitions between asymptotic states, this `in-in' formulation of QFT, as in the CTP formalism, is indispensable. This is certainly true  for the description of non-equilibrium processes  such as  quantum transport in quantum many-body systems.  Powerful functional techniques have been developed to deal with such problems. Indeed, in this paper we  demonstrate a connection between those techniques and our account of QFT measurements.

A crucial  challenge in incorporating quantum informational concepts in QFT is the lack of a common mathematical language. QIT  focuses on the Hilbert space aspects of quantum theory, for example, the geometry of the space of quantum states, the characterization of entanglement and the identification of   physically permissible state transformations. It relies crucially on the way quantum theory describes the extraction of information through measurements.
In contrast, QFT is commonly described through functional methods that lead to the evaluation of unequal-time field correlation functions. The latter contain all information about the system, and they can be easily manipulated  to derive quantities like perturbative scattering amplitudes, expectations of stress-energy tensors and single-particle Wigner functions.

 The mathematical difference is not only technical, as it pertains  to   fundamental notions like locality and causality. In QIT, these issues are  usually expressed  through the  Local Operations and Classical Communication (LOCC) paradigm \cite{LOCC}: The Hilbert space of a quantum informational system is split up as a tensor product
$\otimes_i {\cal H}_i$, where ${\cal H}_i$ is the Hilbert space of the $i$-th subsystem. A local operation on the $i$-th subsystem is a set of completely positive maps ${\cal C}^{(i)}(a)$  on states of ${\cal H}_i$, such that $\sum_a {\cal C}^{(i)}(a) = \hat{I}$, where $a$ denotes  measurement outcomes. Then, the causal structure of the system involves the notion of  classical communication. An operation ${\cal C}^{(i)}(a)$ on a subsystem $i$ may depend on the outcome $b$ of an operation ${\cal D}^{(j)}(b)$ on a subsystem $j$, if the outcome can be communicated to $i$ through a classical channel prior to the operation ${\cal C}^{(i)}(a)$.  QIT usually does not deal with
real-time quantum signal propagation between disconnected subsystems.

In contrast, QFTs focus on spacetime propagation of information, as expressed in the theory's dynamics and spacetime symmetries. These properties are nicely captured by the unequal-time correlation functions, which can be chosen to be manifestly spacetime covariant. Causality is implemented by the requirement that  specific correlation functions  vanish if their arguments are spacelike separated\footnote{To be precise, causality is incorporated in the {\em cluster decomposition} property  \cite{Weinberg}: there is  specific hierarchy of correlation functions $G_n(x_1, x_2, \ldots, x_n)$, where $n = 0, 1, 2, \ldots$ such that $G_{n+m}(x_1, \ldots, x_n, x'_1, \ldots, x'_m) =  G_n(x_1, x_2, \ldots, x_n) G_n(x'_1, x'_2, \ldots, x'_m)$,
 if the cluster of events  $x_1, \ldots, x_n$ is spacelike separated from the cluster $x'_1, \ldots, x'_m$. }.

\subsection{The Quantum Temporal Probabilities Approach: key points}

The key features of the QTP approach to measurements on quantum fields are the following.   First, the apparatus is fully incorporated into the quantum description and it is also treated via QFT. The interaction between the measured system and the apparatus local and causal, in the sense that it is governed by an interaction Hamiltonian that is a local functional of quantum fields.

The measurement apparatus is a macroscopic system that exhibits classical behavior. We describe this behavior through the decoherent histories approach to emergent classicality \cite{Omn1, GeHa2}.  We assume that the pointer variable is a highly coarse-grained observable, so that histories for measurement outcomes satisfy appropriate decoherence conditions.

QTP treats measurements events as localized in space and in time, as is the case for all physical measurement outcomes. For example, consider a solid-state detector that is elementary in the sense that can record only a single event. The detector has a fixed location in a lab, and it records an event at a moment of time that  is determined with finite accuracy. In principle, both position and time can be random variables.  When directing a single particle towards an array of elementary detectors, both the specific  detector that records the particle (i.e., the locus of the record) and the time of recording vary from one run of the experiments to the other. Hence, physical predictions are expressed in terms of probability densities
        \bey
        P(x_1, q_1; x_2, q_2, \ldots, x_n, q_n), \label{probdengen}
        \eey
for multiple detection events. In Eq. (\ref{probdengen}), $x_i$ stand for spacetime points, $q_i$ stand for any other recorded observable and $P$ is a probability density with respect to both $x_i$ and $q_i$.

Since there is no self-adjoint operator for time, there are no ideal (i.e., projective) measurements for time. It follows that none of the probabilities densities (\ref{probdengen}) corresponds to an ideal measurement, not even ones that involve single-measurement event. The probabilities (\ref{probdengen}) are defined through  Positive Operator Valued Measures (POVMs).

 \subsection{Detection probability for a single detector}
Consider a QFT on Minkowski spacetime $M$, with Heisenberg-picture fields $\hat{\phi}_r(x)$ defined on a Hilbert space ${\cal F}$
that carries a unitary representation of the Poincar\'e group.  The index $r$  runs over spacetime and internal indices.

We denote the Hilbert space associated to an apparatus by   ${\cal K}$. We assume that the apparatus follows a world tube ${\cal W} = \R \times U$ in Minkowski spacetime, for some $U \subset \R^3$.
We assume that the size of the apparatus is   much larger than the scale of microscopic dynamics. We will elaborate on the properties of the apparatus later.

We assume a field-apparatus coupling with support in a small spacetime region around a point $x$. The finite spacetime extent of the interaction  mimics the effect of a detection record localized at $x$. Working in the interaction picture, we express the coupling term as
\bey
\hat{V}_x = \int F_x(y) \hat{C}_a(y) \otimes \hat{J}^a(y), \label{VX}
\eey
where $\hat{C}_a(x)$ is a composite operator on ${\cal F}$ that is local with respect to the field $\hat{\phi}_r(x)$ and $a$ runs over spacetime and internal indices. The current operators $\hat{J}^a(x)$ are defined on ${\cal K}$.
 The switching functions $F_x(y)$ are dimensionless. They vanish outside the interaction region and they depend on the motion of the apparatus. The spacetime volume  associated to a switching function is $\upsilon = \int dY F^2_x(y)$.

The problem with the interaction term (\ref{VX}) is that it is not Poincar\'e covariant, because of the presence of the switching function. Indeed, physical interactions are not switched on and off in measurements. The switching function originates from von Neumann's modeling of measurements \cite{vN}, and its role is to localize the system-apparatus interaction in spacetime. Since the switching function is {\em a priori} fixed, time cannot be an random variable, in apparent contradiction to  statements in Sec. 2.2. Indeed, the QTP method, does not require a switching function, and it treats time explicitly as a random variable. However, the resulting expressions for the probabilities are the same to leading order in perturbation theory, modulo some assumptions that are explicitly stated in Sec. 2.4. Since the treatment with the switching functions employs more familiar techniques and it is computationally easier, we chose to employ this for the construction of probabilities in the main text. For completeness, we present the QTP derivation of the same probability formula in the Appendix. The difference between the two derivations is not be important in the context of the QIT applications considered here, but certainly it is important for any  discussion of the relativistic quantum measurement problem, and related foundational issues.

The S-matrix associated to Eq. (\ref{VX}) is $\hat{S}_x = {\cal T} \exp[ - i \int d^4y F_x(y) \hat{C}_a(y) \otimes \hat{J}^a(y)]$, where ${\cal T}$ stands for time ordering. To leading order in the interaction,
\bey
\hat{S}_x = \hat{I} - i \hat{V}_x.
\eey
Let the initial state of the system  be $|\psi\rangle \in {\cal F}$ and the initial state of the apparatus  be $|\Omega\rangle$. A particle record appears if the detector transitions from $|\Omega\rangle$ to its complementary subspace ${\cal K}'$. Furthermore, on ${\cal K}'$, we  measure a property of the particle through a pointer observable $q$. The latter is described by a family of positive operators $\hat{\Pi}(q)$, such that $\sum_{q} \hat{\Pi}(q) = \hat{I} - |\Omega \rangle \langle \Omega|$. The pointer observable is typically very coarse, and we assume that it is stationary under spacetime translations by the self-dynamics of the detector, so that the record is preserved after the end of the measurement.

Then, we evaluate  the probability
\bey
\mbox{Prob}(x, q) = \langle \psi, \Omega| \hat{S}^{\dagger}_x[\hat{I} \otimes \hat{\Pi}(q)]\hat{S}_x|\psi, \Omega\rangle
\eey
 that the detector is excited and records a value $q$ to leading order in perturbation theory
\bey
\mbox{Prob}(x, q) = \int d^4y_1 d^4y_2  F_x(y_1) F_x(y_2) G_{ab}(y_1, y_2) \langle \Omega|\hat{J}^a(y_1) \hat{\Pi}(q) \hat{J}^b(y_2)|\Omega\rangle, \label{probX}
\eey
where
\bey
G_{ab}(x, x')  = \langle \psi|\hat{C}_a(x) \hat{C}_b(x')|\psi\rangle,
\eey
is a correlation function for the composite operator.

The probability $\mbox{Prob}(x, q)$ of Eq. (\ref{probX}) is not a density with respect to $x$, because $x$ appears as a parameter of the switching function.   We define an unnormalized probability density $W(x, q)$ with respect to $x$ by dividing $\mbox{Prob}(x, q)$ with the effective spacetime volume $\upsilon$,
\bey
W(x, q) = \upsilon^{-1}\mbox{Prob}(x, q). \label{Prob0b}
\eey

To further proceed in our analysis, we have to make specific assumptions about the detector. First, we assume that the detector carries a representation of the spacetime translation group with generators $\hat{p}^{\mu}$.  At a fundamental level,  the detector Hilbert space also carries a representation of the Poincar\'e group. However, the state $|\Omega\rangle$ is not the Poincar\'e invariant vacuum\footnote{ An actual detector involves a macroscopically large number of fermions---to be precise, it is described by the subspace of the fermionic fields for leptons and baryons that corresponds to macroscopically large, constant, values of the leptonic and baryonic quantum numbers.}; it defines a preferred reference system at which its center of momentum has zero three-momentum. The fact that the apparatus cannot be described by the vacuum breaks the covariance of the measurement model.

We choose a reference point $x_0 $ in  the detector's world-tube, and we write
\bey
\hat{J}^{a}(y) = e^{-i \hat{p} \cdot (y - x_0)} \hat{J}^a(x_0) e^{i \hat{p} \cdot (y - x_0)}.
\eey
It is convenient to take
$|\Omega\rangle$ to be   {\em approximately translation invariant}. Intuitively, this corresponds to the idea that the apparatus is prepared in an initial state that is homogeneous at the length scales that correspond to position sampling and approximately static at the time scales that correspond to time sampling\footnote{In non-relativistic von Neumann measurements, one often takes the initial state of  the detector to be the ground state of the detector Hamiltonian, so it is exactly static. If the initial state is only an approximate eigenstate, its time evolution is usually viewed as part of the measurement's noise.}. In the present context, approximate translation invariance is the requirement that
\bey
\int d^4 x F_{x}(x') \hat{J}^a(x') |\Omega\rangle \simeq \int d^4 x F_{x}(x') e^{-i \hat{p} \cdot (x' - x_0)} \hat{J}^a(x_0)|\Omega\rangle.
\eey
With this assumption, we can write $\langle \Omega|\hat{J}^a(y_1) \hat{\Pi}(q) \hat{J}^b(y_2)|\Omega\rangle = R^{ab}(y_2 - y_1, q)$,
 where
\bey
R^{ab}(x, q) := \langle \Omega| \hat{J}^a(x_0) \sqrt{\hat{\Pi}}(q) e^{-i\hat{p}\cdot x}\sqrt{\hat{\Pi}}(q)\hat{J}^b(x_0)|\Omega\rangle \label{detkern}
\eey
We will refer to $R^{ab}(y, q)$ as the {\em detector kernel}.

The simplest choice for the switching functions $F_x$ are Gaussians,
\bey
F_x(y) = \exp[ - \frac{1}{2} D(x, y)],
\eey
where $D$ a Euclidean distance function on Minkowski spacetime. Such functions are defined in terms of a dimensionless Euclidean metric $g^{(E)}_{\mu \nu}$ on Minkowski spacetime. A simple and physically relevant metric is determined by the timelike vector field $u_{\mu}$ normal along the world-tube, by
\bey
g_{\mu \nu} = \frac{1}{\delta_t^2} u_{\mu} u_{\nu} + \frac{1}{\delta_x^2} ( u_{\mu} u_{\nu}  + \eta_{\mu \nu}).
\eey
where $\delta_t$ is the  temporal accuracy and $\delta_x$ is the special accuracy of the detector. As these quantities correspond to the sampling of the detection event, they are both macroscopic scales.

 Gaussian switching functions satisfy the identity
\bey
f(x) f(x') = f^2\left(\frac{x+x'}{2}\right) \sqrt{f}(x - x'). \label{gauidty}
\eey
The spacetime volume $\upsilon$ of the interaction region  is $\upsilon =\pi^2 \delta_t \delta_x^3$. We note that the function   $\sigma(x): = \frac{1}{\upsilon} f^2(x)$ is a normalized probability density on $M$. Then, we write
\bey
W(x, q) = \int d^4x'  \sigma(x - x') P(x', q), \label{probX20}
\eey
where
\bey
P(x, q) = \int d^4y  \sqrt{f}(y) R^{ab}(y, q) G_{ab}(x - \frac{1}{2}y, x +\frac{1}{2}y), \label{probX2}
 \eey


The probability distribution $W(x, q) $ is the convolution of $P(x, q)$ with the probability density $\sigma(x)$ that incorporates the accuracy of our measurements. If $P(x, q)$ is non-negative and the scale of variation in $x$ is much larger than both $\delta_t$ and $\delta_x$, we can treat $P(x, q)$  as a finer-grained version of $W(x, q)$ and employ this as our probability density for detection.

The kernel $R^{ab}(x, q)$ is typically characterized by some correlation length-scale $\ell$ and some correlation time-scale $\tau$, such that $ R^{ab}(x, q)  \simeq 0$ if $|t(\xi)| \gg \tau $ or  $|{\bf x}(\xi)|\gg \ell$.
Both scales $\ell$ and $\tau$ are microscopic and characterize the constituents of the apparatus and their dynamics. If $\ell \ll \delta_x$ and $\tau \ll \delta_t$, then $R^{ab}(x, q) \sqrt{f}(x) \simeq R^{ab}(x, q)$ and we obtain an expression for the probability density  $P(x, q)$ that is sampling-independent
\bey
P(x, q) = \int d^4 y     R^{ab}(y, q) G_{ab}(x - \frac{1}{2}y, x +\frac{1}{2}y). \label{prob1aa}
\eey
The probability densities (\ref{prob1aa}) are not normalized to unity. In general, the total probability of detection $P_{det} = \sum_{q} \int_{\cal W} d^4x  P(q, x) $ must be a small number, for perturbation theory to be applicable. There is always a probability $P(\emptyset) = 1 - P_{det}$ of no detection. We normalize probabilities by dividing $P(x, q)/P_{det}$, i.e., by conditioning the probability densities $P(q, x)$ with respect to the existence of a detection record.

\subsection{Remarks}

\noindent 1. The  definition (\ref{Prob0b}) of a spacetime density with respect to time is fully justified in classical probability theory, but it is not rigorous for quantum probabilities. There reason is that it involves a combination of probabilities defined with respect to different experimental set-ups, i.e., different switching functions for the Hamiltonians.
Nonetheless,   Eq. (\ref{Prob0b}) can  be derived as a genuine probability density in the context of the QTP method \cite{QTP3, QTPdet}, to leading order in the field-apparatus coupling. This derivation is reproduced in the Appendix.

\smallskip

\noindent 2. Probabilities are defined here using the Born rule for the pointer variable. Note that, strictly speaking, the probabilities  in  von Neumann measurements are defined at a time {\em after} the function has been switched off, and not at the time when the switching is on.
The  derivation in the appendix employs a  probability assignment for histories which incorporates both the Born rule and the state reduction rule.

\smallskip

\noindent 3.    In the proper QTP derivation of detection probabilities, the interaction is present at all times, as the total description must be time-translation invariant.
The smearing functions $F_x(y)$ are not interpreted in terms of a switching-on of the interaction,
 but they describe the {\em sampling} of the spacetime point. Hence, the spacetime volume $\hat{\upsilon}$ is a measure of {\em coarse-graining}, i.e., of the inaccuracy in the determination of the spacetime point. This point is important for a proper derivation, because probabilities can only be defined   for histories that satisfy a decoherence condition, for which coarse-graining is a prerequisite. In principle, a preferred value of $\upsilon$ that corresponds to the coarse-graining scale at which probabilities are well-defined is determined from first-principles---see \cite{QTP1} for explicit calculations in simple models.
  This means that not all sampling functions $F_x(y)$ are acceptable:  their support  cannot be made arbitrarily small. Such constraints cannot be seen in the derivation that we presented here.

\smallskip

\noindent 4. QTP leads to different predictions beyond the lowest order in the system-apparatus interaction. However, in many set-ups only the lowest order  terms can be said to provide a meaningful signal, i.e., a correlation between  observables in the measured field and pointer variables in the apparatus. Higher order interaction terms often degrade such correlations, and for this reason they can be conceptualized as noise. This  is the case, for example, in Glauber's photodetection theory.

\smallskip

\noindent 5. In general, the field-apparatus couplings  are fixed by the standard model of particle physics. Even in idealized models for detectors, we have to specify the physical process through which detection takes place. This information is encoded in  the composite operators $\hat{C}(x)$.
Consider, for example, the case of a QFT with a single scalar field $\hat{\phi}(x)$. The choice $\hat{C}(x) = \hat{\phi}(x)$ corresponds to detection of particles through absorption. The choice $\hat{C}(x) = :\hat{\phi}^2(x):$ corresponds to particle detection through scattering, through terms that involve one creation and one annihilation operator. This observation generalizes to fields of arbitrary spin.

\subsection{Detection probability for multiple detectors}
We proceed to the derivation of a probability formula for the case  that the field interacts with $n$ detectors. Each detector corresponds to a Hilbert space ${\cal K}_i$, $ i = 1, 2, \ldots, n$. The total Hilbert space of the system is ${\cal F} \otimes {\cal K}_1 \otimes {\cal K}_2 \otimes \ldots \otimes {\cal K}_n$. The coupling operator of the $i$-th detector is
\bey
\hat{V}^{(i)}_x = \int d^4y F_x(y) \hat{C}^{(i)}_a(y) \otimes \hat{I} \otimes \ldots \hat{J}_{(i)}^a(y) \otimes \ldots \otimes \hat{I}, \label{VXn}
\eey
where $\hat{J}_{(i)}^a(y)$ is a current operator for the $i$-th detector and $\hat{C}_a^{(i)}$ is the associated composite operator.

 We denote the S-matrix for the $i$-th detector by $\hat{S}^{(i)}_x$. Again, to leading order in the interaction $\hat{S}^{(i)}_x = \hat{I} - i \hat{V}^{(i)}_x $.
 We also denote the initial state of the $i$-th detector by $|\Omega_i\rangle$, and the measurement operators as $\hat{\Pi}^{(i)}(q_i)$.

The key point is that the S matrix for the total interaction of the field  is defined by time ordering with respect to the spacetime points  $x_1, x_2, \ldots, x_n$ associated to the detectors,
\bey
\hat{S}_{x_1, x_2, \ldots, x_n} = {\cal T}\left[ \hat{S}^{(1)}_{x_1} \hat{S}^{(2)}_{x_2} \ldots \hat{S}^{(n)}_{x_n}   \right];
\eey
here, ${\cal T}$ stands for time ordering.

To leading order in perturbation theory,  the probability density for $n$ measurement events is
\bey
W_n(x_1, q_1; x_2, q_2; \ldots; x_n, q_n) = \int d^4x'_1 \ldots d^4 x'_n \sigma^{(1)}(x_1 - x'_1) \ldots \sigma^{(n)}(x_n - x'_n)\nonumber \\ P(x'_1, q_1;  x'_2, q_2; \ldots; x'_n, q_n)
\eey
where
\bey
P_n(x_1, q_1; x_2, q_2; \ldots; x_n, q_n) = \int d^4 y_1 \ldots d^4 y_n   \sqrt{f^{(1)}}(y_1) \ldots \sqrt{f^{(n)}}(y_n)
R_{(1)}^{a_1b_1}(y_1, q_1) \ldots
\nonumber \\
\times \ldots R_{(n)}^{a_nb_n}(y_n, q_n)  G_{a_1 \ldots a_n, b_1\ldots b_n}(x_1 - \frac{1}{2}y_1, \ldots, x_n - \frac{1}{2}y_n; x_1 + \frac{1}{2}y_1, \ldots, x_n - \frac{1}{2}y_n). \hspace{0.2cm} \label{probdenN}
\eey
Here $R^{(i)}(x, q)$ is the measurement kernel for the $i$-th detector. The field correlation function $G_{a_1 \ldots a_n, b_1\ldots b_n}(x_1, \ldots, x_n; x_1', \ldots, x_n')$ is given by
\bey
G_{a_1 \ldots a_n, b_1\ldots b_n}(x_1, \ldots, x_n; x_1', \ldots, x_n') = \langle \psi| {\cal T}^*[\hat{C}_{b_1}^{(1)}(x'_1) \ldots \hat{C}_{b_n}^{(n)}(x'_n) ]
\nonumber \\
\times {\cal T} [\hat{C}_{a_n}^{(n)}(x_n) \ldots \hat{C}_{a_1}^{(1)}(x_1)]|\psi\rangle \label{correl}
\eey
where ${\cal T}^*$ stands for reverse time ordering.

Again we note that in the appropriate regime, the probability becomes independent of the switching functions, and equal to
\bey
P_n(x_1,q_1; x_2, q_2; \ldots; x_n, q_n) = \int d^4 y_1 \ldots d^4 y_n     R_{(1)}^{a_1b_1}(y_1, q_1) \ldots R_{(n)}^{a_n1b_n}(y_n, q_n)
\nonumber \\
\times G_{a_1 \ldots a_n, b_1\ldots b_n}(x_1 - \frac{1}{2}y_1, \ldots, x_n - \frac{1}{2}y_n; x_1 + \frac{1}{2}y_1, \ldots, x_n + \frac{1}{2}y_n) \label{probden4}
\eey
\subsection{The detector kernel}
The contribution of each detector to the probability is determined by the detector kernel $R^{ab}(x, q)$, defined by Eq. (\ref{detkern}). The detector kernel coincides with the matrix elements of a unitary operator,
\bey
R^{ab}(x, q) = \langle a, q|e^{i \hat{p} \cdot (x - x_0)}|b, q\rangle,
\eey
where $|a, q\rangle = \sqrt{\hat{\Pi}}(q) \hat{J}^a(x_0)|\Omega\rangle$.  If $\hat{\Pi}$ is a projector, then vectors $|a, q\rangle$ with different values of $q$ are orthogonal: $\langle a, q| b, q'\rangle = 0$ for $q \neq q'$.

The Fourier transform of the detector kernel
\bey
\tilde{R}^{ab}(\xi, q) = \int d^4x e^{-i \xi\cdot x} R^{ab}(x, q),
\eey
is given by
\bey
\tilde{R}^{ab}(\xi, q) = (2\pi)^4 e^{i \xi\cdot x_0} \langle a, q|\hat{E}_{\xi}|b, q\rangle,
\eey
where $\hat{E}_{\xi} = \delta^4(\hat{p} - \xi)$ is the projector onto the subspace with four-momentum $\xi^{\mu}$,  $\langle a, q|\hat{E}_{\xi}|a, q\rangle \geq 0$.
The momentum four vector associated to the detector is timelike and the associated energy $p^0$ is positive. This means that $\tilde{R}^{ab}(p, q) = 0$ for spacelike $p$, or if $p^0 < 0$.

For a  scalar composite operator $\hat{C}(x)$, we can drop the indices $a, b, \ldots$ from the detector kernel and write simply $R(x, q)$. If we record no other observable, we simply write $R(x) := \langle \omega|e^{i \hat{p} \cdot (x - x_0)}|\omega\rangle$, where $|\omega\rangle =  \hat{J}^a(x_0)|\Omega\rangle$. Hence, the detector kernel is determined by the energy-momentum distribution of the `threshold' state $\omega$.


\subsection{Index notation for probabilities}

It is convenient to express the probability densities (\ref{probden4}) using an abstract notation. We use small Greek indices $\alpha, \beta, \gamma \ldots$ for the pairs  $(x, a)$ where $x$ is a spacetime point and $a$ the internal index for the composite operators $\hat{C}_a$. All indices  in a time-ordered product are upper, and all indices in an anti-time-ordered product are lower. Hence, we write the correlation functions (\ref{correl}) as
\bey
G^{\alpha_1 \alpha_2 \ldots \alpha_n}_{\beta_1 \beta_2 \ldots \beta_n} \nonumber
\eey

We denote by $z$ the  pairs $(x, q)$, where $x$ is a spacetime point and $q$ any other recorded observable or the event $\emptyset$ of no detection. If we denote by $\Gamma$ the set of possible values for $q$, then $z$ takes values on the set $Z:= M \times \Gamma \cup \{\emptyset\}$.
We will write the kernel
\bey
\sigma[x - \frac{1}{2}(y+y')] \sqrt{f}(y-y') R^{ab}(y - y', q) \nonumber
\eey
as $R_\alpha^\beta(z)$ where $\alpha$ stands for $(y, a)$, $\beta$ for $(y', b)$ and $z$ for $(x, q)$. We will use the same symbol for the approximate expression $\delta[x - \frac{1}{2}(y+y')]R^{ab}(y - y', q)$. We use the Einstein summation convention over Greek indices, to denote sum over the discrete index $a$ and spacetime integral.

Using the index notation,  we write the probability formula (\ref{probden4}) as
\bey
P_n(z_1, z_2, \ldots, z_n)  = G^{\alpha_1 \alpha_2 \ldots \alpha_n}_{\beta_1 \beta_2 \ldots \beta_n} \;\; {}^{(1)} R_{\alpha_1}^{\beta_1}(z_1) \; {}^{(2)} R_{\alpha_2}^{\beta_2}(z_2) \ldots \; {}^{(n)} R_{\alpha_n}^{\beta_n}(z_n). \label{probdenr}
\eey
There are two ways of viewing Eq. (\ref{probdenr}). On one hand, it corresponds to a POVM defined on the Hilbert space ${\cal F}$ of the quantum field, with alternatives in the set $Z^n$. On the other hand, we can consider a vector space ${\cal S}$ of smearing functions $f_a$ to the composite operators $\hat{C}^a(x)$, so that $\hat{C}(f) = \int d^4 x \hat{C}^a(x) f_a(x)$. The detector kernel is an element of ${\cal S} \otimes {\cal S}^*$, where ${\cal S}^*$ stands for the dual vector space of ${\cal S}$. Then, Eq. (\ref{probdenr}) defines a positive linear functional on $({\cal S} \otimes {\cal S}^*)^n$ that corresponds to the correlation functions
$G^{\alpha_1 \alpha_2 \ldots \alpha_n}_{\beta_1 \beta_2 \ldots \beta_n}$

\section{Relation of the QTP approach  to the Closed-Time-Path formalism}

The probability density (\ref{probdenN}) for $n$ measurement events is a linear functional of the $2n$-point unequal-time correlation function (\ref{correl}). This correlation function has
 $n$ time-ordered arguments and $n$ anti-time-ordered arguments.  It does not appear in the usual S-matrix description of QFT; the correlation functions in the S-matrix
 description involve only time-ordered arguments. Rather,  the correlation function (\ref{correl}) appears in the
  Schwinger-Keldysh or Closed-Time-Path (CTP) formalism of QFT.

  In what follows, we will review the motivation and function of the CTP formalism, and then we will show  how it relates to the QTP probability assignment.

\subsection{S-matrix vs CTP correlation functions}
By the Dyson formula, the S-matrix for any QFT involves Feynman (i.e., time-ordered) correlation functions of local composite operators $\hat{C}(x)$ of the fields.
We can  generate the time-ordered correlation functions of a composite operator  from the variational derivatives of a generating functional $Z[J] = \langle 0|\hat{U}[J]|0\rangle$, where
\bey
\hat{U}[J] = {\cal T} \exp\left[i \int d^4 X J^a(X)\hat{C}_a(X)\right]. \label{source}
\eey
 The generating functional  admits the path
integral representation
\begin{equation}
\label{R1}Z[J] =\int~D\phi~e^{i \{S[\Phi]+\int d^4xJ(x)C (x)\}},
\end{equation}
where $S[\phi]$ is the classical action functional and $C(x)$ is the classical counterpart of the composite operator $\hat{C}(x)$.

It is convenient to work with the generating functional $W[J] = - i \ln Z[J]$. In perturbation theory, the coefficients of the series expansion of $W$
 with respect to $J$ are the
connected Feynman graphs. In particular,
we may introduce the `background' or `mean' field
\begin{equation}
\label{R2}C_a (x)={\frac{\delta W[J]}{\delta J^a(x)}}
\end{equation}
The  Legendre transformation of $W$ provides a more efficient representation of the correlation functions
\begin{equation}
\label{R3}\Gamma [C ]=W[J]-\int d^4x~J^a(x)C_a (x)
\end{equation}
since the series  expansion of $\Gamma $ involves only one-particle
irreducible   Feynman graphs. Despite the formal similarity, the inverse of (\ref{R2}),
\begin{equation}
\label{R5}{\frac{\delta\Gamma}{\delta C_a (x)}}=-J^a(x).
\end{equation}
is not an equation of motion for the mean field, because it is complex-valued even for hermitian composite operators. This is due to the fact that the  `mean' field is not a true expectation value, but
rather a matrix element of the composite operator between {\em in} and {\em out} states
\cite{hartlehu}.

To define a proper mean-field equation of motion we adopt the  Closed-Time-Path formalism of Schwinger \cite{ctp1} and Keldysh\cite{ctp2}.  To this end, we couple the field to two different external sources $J^a(x)$ and $\bar{J}^a(x)$, and we define the CTP generating functional
\bey
Z_{CTP}[J, \bar{J}] = \langle \psi_0|\hat{U}^{\dagger}[\bar{J}]\hat{U}[J]|\psi_0\rangle,
\eey
By definition, $Z[J, J] = 1$ and $Z^*[J, \bar{J}] = Z[\bar{J}, J]$. The state $|\psi_0\rangle $ is defined in the distant past, i.e., prior to any time at which
 $J(x)$ has support---it is an {\em in} state. Note that a large class of initial states, relevant to particle experiments,  can be reconstructed by the action of source terms of the form (\ref{source}). Hence, for a large class of problems, the generating function $Z_{CTP}$ for the QFT vacuum $|0\rangle$ contains all physically relevant information.

The CTP generating functional describes correlation functions with $n$ time-ordered and $m$ anti-time-ordered entries,
\bey
G^{n,m}_{a_1 \ldots a_m, b_1\ldots b_n}(x_1, \ldots, x_n; x_1', \ldots, x_n') = \langle \psi_0| {\cal T}^*[\hat{C}_{b_1}^{(1)}(x'_1) \ldots \hat{C}_{b_m}^{(n)}(x'_m) ]
\nonumber \\
\times {\cal T} [\hat{C}_{a_n}^{(i)}(x_n) \ldots \hat{C}_{a_1}^{(1)}(x_1)]|\psi_0\rangle \label{correl8}
\eey
as functional derivatives of $Z_{CTP}[J, \bar{J}]$,
\bey
G^{n,m}_{a_1 \ldots a_m, b_1\ldots b_n}(x_1, \ldots, x_n; x_1', \ldots, x_n') = i^{n-m} \left(\frac{\partial^{n+m}Z_{CTP}}{\delta J^a_1(x_1) \ldots \delta J^{a_n}(x_n)\delta \bar{J}^b_1(x'_1) \ldots \delta \bar{J}^{b_m}(x'_n) }\right)_{J = \bar{J} = 0}. \nonumber
\eey
 For a vacuum initial state, the generating functional has a path integral expression
\begin{equation}
\label{R8}  Z_{CTP}[J, \bar{J}]
=\int~D\phi~D\bar{\phi} \; e^{i\{S[\phi]-S[\bar{\phi}]+\int d^4x~[J^a(x)C_a (x)
-\bar{J}(x)\bar{C}_a (x)]\} },
\end{equation}
where $\bar{C}$ is defined a functional of $\bar{\phi}$.
Thus we formally double the degrees of freedom and
we integrate over pairs of histories $(\phi,\bar{\phi})$ that vanish in the asymptotic past and future.

We can actually define two background fields $C_a(x)$ and $\bar{C}_a(x)$ by taking variations with
respect to both sources,
\bey
C_a(x)  = \frac{\delta W}{\delta J^a(x)}  \hspace{1cm} \bar{C}_a(x)  = -\frac{\delta W}{\delta \bar{J}^a(x)}
\eey
where $W[J, \bar{J}] = -i \ln Z_{CTP}[J, \bar{J}]$.
The closed time-path effective action  is defined as the double Legendre
transform of $W$
\begin{equation}
\label{R11}\Gamma [C, \bar{C}]= W[J, \bar{J}] - \int d^4 x J^a(x) C_a(x) +  \int d^4 x \bar{J}^a(x) \bar{C}_a(x).
\end{equation}
The equations of motion now form a coupled system
\begin{equation}
\label{R12}\frac{\delta\Gamma}{\delta C_a (x)}= J_a(x) \hspace{1cm} \frac{\delta\Gamma}{\delta \bar{C}_a (x)}= \bar{J}_a(x)
\end{equation}
These equations coincide
when $J=J'$ is the
physical external source, and their solution $C = \bar{C}$
is the physical mean field. Of course, both background fields are
identified only after the computation of the  variational derivative \cite{ch87}.
Unlike the in-out formulation, the equations of motion (\ref{R12})
are always real and causal. The cases where the in-out effective
action becomes complex correspond to dissipative evolution for the mean field. The doubling of degrees of freedom is essential for the description of dissipation, because  no action functional of a single  mean field
alone could possibly lead to the right  equations of motion.

The CTP effective action formalism can be used to treat fully non-equilibrium dynamics of open quantum systems,
going far beyond the traditional linear response theory which are confined to near-equilibrium conditions. In this context, we are often interested in calculating mean-field
expectation values $\langle \hat{C}_a(x)\rangle$ that may describe the evolution of classicalized macroscopic field observables.

The CTP effective action satisfies $\Gamma [C , C ] = 0$ and $\Gamma
[C,\bar{C}]=-\Gamma [\bar{C},C]^*$. Therefore, if we separate
the CTP effective action in its real and imaginary parts, the former is an odd
function of $C_a$ and $\bar{C}_a$ and the latter is even. Let   $C_a^{(0)} = \bar{C}_a^{(0)}$ be an extremum of the CTP effective action, which corresponds to a solution
of the mean field equations of motion with zero sources. We take linearized perturbations $\Delta C_a = C_a - C_a^{(0)}$ and $\Delta \bar{C}_a = \bar{C}_a - C_a^{(0)}$. Then, the
CTP effective action must take the form
\bey
 \Gamma [\Delta C, \Delta \bar{C}] ={\frac{1}{2}}\int~d^dxd^dx^{\prime}\{-[%
\Delta C_a(x) ](x){\cal D}^{ab}(x,x^{\prime})
\{\Delta C_b\}(x^{\prime})
\nonumber \\
+i[\Delta C_a](x)N^{ab}(x,x^{\prime}) [\Delta C_b
](x^{\prime})\} \nonumber \\
\eey
where $[\Delta C_a ]=(\Delta C_a-\Delta \bar{C}_a)$, $\{\Delta C_a\}=(\Delta
C_a + \Delta \bar{C}_a)$, and ${\cal D}^{ab}$ and $N^{ab}$ are two non-local kernels. Superficially, $N^{ab}$ appears redundant, because it does
appear in the mean-field equations of motion. The latter take the simple form
\begin{equation}
\int~d^dx^{\prime}{\cal D}^{ab}(x,x^{\prime})\Delta C_b(x^{\prime}) = J^a(x), \label{meanfielddev}
\end{equation}
where $J^a(x)$ is an external source. The kernel $N^{ab}$ contains  information  about the {\em departure}
of the actual evolution from the mean-field equation (\ref{meanfielddev}).  This deviation reflects the fact that the mean field is coupled to a {\em stochastic} external source \cite{FeVe1, FeVe2, CaLe, HPZ1}, rather than a deterministic one like $J^a(x)$ in Eq. (\ref{meanfielddev}). This means that the effective classical equations of motion are of the form
\begin{equation}
\int~d^dx^{\prime}{\cal D}^{ab}(x,x^{\prime})\Delta C_b(x^{\prime}) = J^a(x) + \xi^a(x), \label{meanfielddev2}
\end{equation}
where $\xi^a(x)$ has vanishing expectation---thus justifying Eq. (\ref{meanfielddev})---but non-zero correlator
$\langle \xi(x)^a \xi^b(x')\rangle = N^{ab}(x, x')$. This is why $N$ is referred to as a the {\em noise kernel}. Hence, CTP allows us to explore how higher-order correlation functions
are manifested as noise in the mean-field (classical) equations of motion.

\subsection{Measurements and the CTP generating functional}

The relation between the QTP description of measurements and the CTP formalism is more transparent, if we use the index notation of Sec. 2.7. For consistency, the sources $J$ have a lower Greek index, and the sources $\bar{J}$ have an upper Greek index. Then, we write the CTP generating functional as
\bey
Z_{CTP}[J, \bar{J}] = \sum_{n, m = 0}^{\infty} \frac{ i^{m-n}}{n!m!}  G^{\alpha_1 \ldots \alpha_n}_{\beta_1\ldots \beta_m} J_{\alpha_1} \ldots J_{\alpha_n} \bar{J}^{\beta_1} \ldots \bar{J}^{\beta_m},
\eey
or, conversely,
\bey
G^{\alpha_1 \ldots \alpha_n}_{\beta_1\ldots \beta_m}  = i^{n-m} \left(\frac{\partial^{n+m}Z_{CTP}[J, \bar{J}]}{\partial J_{\alpha_1} \ldots \partial J_{\alpha_n} \partial \bar{J}^{\beta_1} \ldots \partial \bar{J}^{\beta_n}}\right)_{J = \bar{J}+0}.
\eey
The probability densities (\ref{probdenr}) involve  {\em balanced} correlation functions, i.e., correlation functions  with an equal number of upper and lower indices. We can construct a generating functional that contains only such functions. The key observation is that such correlations contribute to the sum only through products of the form $J_\alpha \bar{J}^{\beta}$. Hence, the natural source for a diagonal generating functional $Z^d_{CTP}$ that only involves balanced correlation functions is a `tensor' $L_{\alpha}^{\beta}$. We define
\bey
Z^d_{CTP}[L] = \sum_{n=0}^{\infty} \frac{1}{n!}G^{\alpha_1 \ldots \alpha_n}_{\beta_1\ldots \beta_m} L_{\alpha_1}^{\beta_1} \ldots L_{\alpha_N}^{\beta_N}.
\eey

Suppose now that we consider only measurements of a single type, i.e., all detector kernels $R_{\alpha}^{\beta}(z)$ are identical. Then, we can define a moment-generating functional for all probability densities
 (\ref{probdenr}), in terms of sources $j(z)$,
\bey
Z_{QTP}[j] = \sum_{n=0}^{\infty} \sum_{z_1, z_2, \ldots, z_n} \frac{1}{n!}  P_n(z_1, z_2, \ldots, z_n) j(z_1) \ldots j(z_n). \label{zqtp}
\eey
It is straightforward to show that
\bey
Z_{QTP}[j]  = Z^d_{CTP}[R\cdot j], \label{fundamental}
\eey
where $(R\cdot j)_A^B = \sum_{z} R_A^B(z) J(z)$.

Eq. (\ref{fundamental}) is a fundamental relation for quantum measurements in QFT, as it relates the moment generating functional for a hierarchy of measured probability densities to the generating functional of unequal-time correlation functions

It is straightforward to write a path integral expression for $Z^d_{CTP}[L]$ for a vacuum initial state
\bey
Z^d_{CTP}[L] = \int~D\phi~D\bar{\phi} \; e^{iS[\phi]- i S[\bar{\phi}]+\int d^4x d^4x' C_a(x) \bar{C}_b(x') L^{ab}(x, x') }.
\eey
To obtain a simple path integral expression for a broader class of states, we recall that many field initial states can be obtained from the action of an external source $\zeta(x)$ on the vacuum, i.e., they are of the form $|\psi_0\rangle = \hat{U}[\zeta]|0\rangle$, where now we write $\hat{U}[\zeta] = {\cal T} \exp\left[i \int d^4 X \zeta^k(X)\hat{A}_k(X)\right]$ in terms of composite operators $\hat{A}_k(x)$ that differ, in general from $\hat{C}_a(x)$.

To see this, consider the case of a free scalar field $\hat{\phi}(x)$. Choosing $\hat{A}(x) = \hat{\phi}(x)$, $\hat{U}[\zeta]$ is a Weyl operator, and $|\psi_0\rangle$ is a field coherent state. For single-event measurements, and for any state with a fixed number $N$ of particles,
the probability density (\ref{prob1aa}) depends only on the one-particle reduced density matrix $\rho_1({\bf k}, {\bf k'}) = N^{-1} \langle \psi_0|\hat{a}^{\dagger}_{\bf k} \hat{a}_{{\bf k}'}|\psi_0\rangle$, where $\hat{a}_{\bf k}$ and $\hat{a}^{\dagger}_{\bf k}$ are the creation and annihilation operators of the field. It is straightforward to show that one can reproduce any pure $\rho_1$ with an appropriate choice of $\zeta(x)$.
 For $\hat{A}(x)  =  :\hat{\phi}(x)^2:$, $\hat{U}[\zeta]$ is a product of a  Bogoliubov transformation with a Weyl operator, and, hence, $|\psi_0\rangle$ is a squeezed state. We can reproduce a large class of two-particle reduced density matrices $\rho_2({\bf k}_1, {\bf k}_2;{\bf k'}_1, {\bf k}'_2) = N^{-2} \langle \psi_0|\hat{a}^{\dagger}_{{\bf k}_1} \hat{a}^{\dagger}_{{\bf k}_2}\hat{a}_{{\bf k}'_2} \hat{a}_{{\bf k}'_1}|\psi_0\rangle$, including entangled ones,  with appropriate choices of $\zeta$.

Hence, for a quantum state that is obtained from an external source $\zeta_k$, we  write the path integral expression
\bey
Z^d_{CTP}[f, L] = \int~D\phi~D\bar{\phi} \; e^{iS[\phi]- i S[\bar{\phi}] + i  \int d^4x~\zeta^k(x)[ A_k (x)
-\bar{A}_k (x)] + \int d^4x d^4x' C_a(x) \bar{C}_b(x') L^{ab}(x, x') },
\eey
where we must assume that the spacetime support of the kernel $L^{ab}$ is later than the support of $\zeta$ (state preparation is prior to measurement).
By Eq. (\ref{fundamental})
\bey
Z_{QTP}[f, j] = \int~D\phi~D\bar{\phi} \; e^{iS[\phi]- i S[\bar{\phi}] + i  \int d^4x~\zeta^k(x)[ A_k (x)
-\bar{A}_k (x)]}
\nonumber \\
\times e^{ \sum_z \int d^4x d^4x' C_a(x) \bar{C}_b(x') R^{ab}(x, x'; z) j(z)} \label{ZQTP}
\eey
The probability densities for $n$ measurement events are obtained from functional variation of $Z_{QTP}[\zeta, j] $ with respect to $j$ at $ j = 0$
For example, the single-event probability density $\hat{P}_1(x, z)$ of Eq. (\ref{prob1aa}) is given by the path integral
\bey
P_1(x, q) =   \int~D\phi~D\bar{\phi} \left(\int d^4y C_a(x+\frac{1}{2}y) \bar{C}_b(x- \frac{1}{2}y) K^{ab}(y, q) \right)
\nonumber \\
     \times         e^{iS[\phi]- i S[\bar{\phi}] + i  \int d^4x~\zeta^k(x)[ A_k (x)
-\bar{A}_k (x)]}. \nonumber
\eey
Expressions such as the above provide an explicit link between concepts of quantum measurement theory like POVMs and  the practical and highly successful  functional language of QFT. We believe that this link is essential for a local and covariant definition of quantum informational notions in QFT.

\section{Applications}
Next, we present three immediate applications of the QTP formalism. First, we review how the QTP formalism leads to the construction of time of arrival probabilities for relativistic particles. Second, we derive Unruh-DeWitt detectors from the QTP probability assignment at a particular limit. Third, we identify the regime in which Glauber's photodetection theory applies.

\subsection{Time-of-arrival probabilities}
The first application of the QTP formalism is the construction of probabilities for the time-of-arrival. The simplest   time-of-arrival measurement involves a particle  prepared on an initial state $|\psi_0 \rangle$ that is localized around $x = 0$ and has positive mean momentum. The issue is to construct the probability
$P(L, t)dt$ that the particle is detected at $x = L$ at some moment between $t$ and $t+\delta t$. The problem is that no unique time-of-arrival probability distribution exists \cite{ML, ToAbooks}. The absence of a self-adjoint time operator    \cite{Pauli} means that we cannot use Born's rule for this task.

A time-of-arrival probability distribution can be constructed from Eq. (\ref{prob1aa}), in absence of any additional observable $q$. The simplest case, corresponding to a free scalar field $\hat{\phi}(x)$ of mass $m$, was analysed in Ref. \cite{QTP3}. For a scalar composite operator $\hat{C}(X)$, we obtain a probability density for detection
\bey
P(x) = \int d^4 \xi    \; R(\xi) G(x - \frac{1}{2}\xi, x +\frac{1}{2}\xi). \label{pxx}
\eey
In a  time-of-arrival measurement, a detector is placed at a macroscopic distance $L$ from the particle source. If $L$ is much larger than the size of the detector, only particles with momenta along the source-detector axis are recorded. Hence, the problem is  reduced to two spacetime dimensions. Hence, we express the spacetime coordinate $x = (t, L)$. Then, Eq. (\ref{pxx}) becomes
\begin{eqnarray}
P(t, L) = \int \frac{dpdp'}{2\pi } \frac{\rho(p,p')}{2\sqrt{\epsilon_p \epsilon_{p'}}} \; \tilde{R}\left( \frac{p+p'}{2}, \frac{\epsilon_p + \epsilon_{p'}}{2}\right) e^{i(p-p')L - i (\epsilon_p - \epsilon_{p'})t}, \label{ptx}
\end{eqnarray}
where $\epsilon_p = \sqrt{p^2+m^2}$, and $\tilde{R}(E, p)$ is the Fourier transform of $R(x) = R(t, L)$.

We normalize Eq. (\ref{ptx}) by treating $L$ as a parameter and $t$ as a random variable. The procedure is described in Ref. \cite{QTP3}. The end result is a probability distribution
\begin{eqnarray}
P(t, L) = \int \frac{dpdp'}{2\pi } \tilde{\rho}(p,p')  \sqrt{v_p v_{p'}} S(p,p') e^{i(p-p')L - i (\epsilon_p - \epsilon_{p'})t}, \label{ptxb}
\end{eqnarray}
where $v_p = p/\epsilon_p$ is the particle velocity.  $S(p, p')$ are the matrix elements $\langle p|\hat{S}|p'\rangle$ of the {\em  localization operator} $\hat{S}$, defined by
\begin{eqnarray}
 \langle p|\hat{S}|p'\rangle  := \frac{\tilde{R}\left( \frac{p+p'}{2}, \frac{\epsilon_p + \epsilon_{p'}}{2}\right)}{\sqrt{\tilde{R}(p, \epsilon_p) \tilde{R}(p', \epsilon_{p'})}}.  \label{lpp}
\end{eqnarray}
By definition, $\langle p|\hat{S}|\hat{p'}\rangle \geq 0 $ and  $S(p, p) = 1$.  Its name originates from the fact that $\hat{S}$ describes the localization of an elementary measurement event.

For a pure state initial state $|\psi\rangle$, Eq. (\ref{ptx}) becomes
\begin{eqnarray}
P(t, L) = \langle \psi|\hat{U}^{\dagger}(t, L) \sqrt{|\hat{v}|}\hat{S}\sqrt{|\hat{v}|}\hat{U}(t, L)|\psi\rangle, \label{ptxc}
\end{eqnarray}
where $\hat{U}(t, L)$ is the  spacetime-translation operator
$\hat{U}(t, L) = e^{i  \hat{p} L - i \hat{H} t} $
and $\hat{v} = \hat{p}\hat{H}^{-1}$ is the velocity operator. Eq. (\ref{ptxc}) defines a positive probability distribution if    $\hat{S}$ is a positive operator, which is always the case if  $\ln \tilde{R}(p, \epsilon_p)$ is  a  convex  function of $p$. Then, the Cauchy-Schwarz inequality applies,
\begin{eqnarray}
\langle p|\hat{S}|p'\rangle \leq \sqrt{\langle p|\hat{S}|p\rangle \langle p'|\hat{S}|p'\rangle} =  1. \label{CSa}
\end{eqnarray}
When Eq. (\ref{CSa}) is saturated, i.e.,  $\langle p|\hat{S}|p'\rangle =  1$, we have maximum localization. This means that $\hat{S} = \delta(\hat{x})$, where $\hat{x} = i \frac{\partial}{\partial p}$.  The corresponding probability density (\ref{ptxc})     was first identified by Le\'on \cite{Leon}, in terms of the eigenvectors of a non-self-adjoint time-of-arrival operator. It generalizes the  non-relativistic time-of-arrival probability distribution of Kijowski \cite{Kij}.

\subsection{The limit of a pointlike detector}
In a pointlike detector, the detector's world-tube shrinks to a single timelike curve $x_0^{\mu}(\tau)$, parameterized by its proper time $\tau$. We implement this limit, by expressing the smearing function as
\bey
F_x(y) = \int d \tau f(\tau - s) \delta^4[y - x_0(\tau)],
\eey
where $f(\tau)$ is a smearing function peaked around $0$, and $s$ is defined by the condition $x_0(s) = x$.

We choose a Gaussian switching function  $f(s) = \exp\left[ - \frac{s ^2}{2\delta_t^2}\right]$.
Then, we  obtain a probability density for detection $W(q, \tau) = \int ds \sigma(\tau - s) P(q, s)$, where $\sigma (s) = (\sqrt{\pi} \delta_t)^{-1}f^2(s)$, and
\bey
P(q, \tau) = \int ds \sqrt{f(s)} G_{ab}[x_0(\tau -\frac{1}{2}s), x_0(\tau+\frac{1}{2}s)]
\nonumber \\
\langle \Omega|\hat{J}^a[x_0(\tau -\frac{1}{2}s)] \Pi(q) \hat{J}^b[x_0(\tau +\frac{1}{2}s)]|\Omega\rangle. \label{pointlike}
\eey
The Unruh-DeWitt detection models \cite{Unruh76, Dewitt} are a special case of pointlike detectors, in which
\bey
\hat{J}^a[x_0(\tau)] = e^{i \hat{h}\tau} \hat{J}^a[x_0(0)]  e^{-i \hat{h}\tau}, \label{UdWc}
\eey
 in terms of a Hamiltonian $\hat{h}$ that is defined with respect to the rest frame of the detector. Such Hamiltonians are well defined if the four-velocity $\dot{x}^{\mu}_0(\tau)$ is a Killing vector of Minkowski spacetime, because then time evolution with respect to $\tau$ corresponds to the time translation of a Lorentzian time coordinate. It is not obvious that such a Hamiltonian exists for non-inertial trajectories. In a general QFT,  Hamiltonians that generate time translations along non inertial time coordinates do not exist \cite{Torre, Helfer}.  However, it may be the case that they can be defined as appropriate limits when the spatial size of the system goes to zero, the vanishing of the system's size canceling the ultraviolet divergences that characterize such Hamiltonians. To the best of our knowledge such an analysis has not been carried out. In this sense, the defining assumption for  Unruh-DeWitt  detectors---for non-inertial trajectories---remains unjustified from first principles.

Assuming the  Unruh-DeWitt  condition (\ref{UdWc}), and taking $|\Omega\rangle$ as the lowest eigenstate of $\hat{h}$, we find
\bey
P(q, \tau) = \int ds \sqrt{f(s)} G_{ab}[x_0(\tau -\frac{1}{2}s), x_0(\tau+\frac{1}{2}s)] \langle \Omega|\hat{\mu}^a \Pi(q) e^{i\hat{h}s} \hat{\mu}^b|\Omega\rangle, \label{pointlike2}
\eey
where we wrote $\hat{\mu}^a =  \hat{J}^a[x_0(0)]$.
If we identify $q$ with the energy $\epsilon$ in the rest frame of the detector, we find that
\bey
P(\epsilon, \tau) = \int ds \sqrt{f(s)} e^{i \epsilon s} G_{ab}[x_0(\tau -\frac{1}{2}s), x_0(\tau+\frac{1}{2}s)] \langle \Omega|\hat{\mu}^a |\epsilon\rangle \langle \epsilon|  \hat{\mu}^b|\Omega\rangle. \label{pointlike3}
\eey
For a scalar coupling operator, where we drop the indices $a, b$ and so on, and for energies such that $\epsilon \delta_t >> 1$, we write
\bey
P(\epsilon, \tau) = |\langle \Omega|\hat{\mu} |\epsilon\rangle|^2 \int ds   e^{i \epsilon s} G_{ab}[x_0(\tau -\frac{1}{2}s), x_0(\tau+\frac{1}{2}s)], \label{pointlike4}
\eey
i.e., we obtain the standard formula for the response of the Unruh-DeWitt detector.

Hence, the probabilities of the Unruh-DeWitt detectors for inertial path can be obtained from the QTP probability formula---that is derived from full QFT---at the limit where the world-tube of the detector shrinks to a single timelike curve.

\subsection{ Derivation of Glauber's theory as a limit}
Consider a detector in inertial motion, i.e., the timelike normal of its world-tube is a time-translation vector field in Minkowski spacetime. For simplicity, we consider a quantum scalar field $\hat{\phi}(x)$, with a composite operators $\hat{C}(x) = \hat{\phi}(x)$. We consider particles with $m = 0$, i.e., scalar photons.

A   coherent state  $|z\rangle$  of the scalar field is defined as an eigenvector of the annihilation operator
\begin{eqnarray}
\hat{a}_{k}|z\rangle = z(k)|z\rangle,
\end{eqnarray}
where $z(k)$ is a square-integrable complex-valued function; here $k = ({\bf k}, \omega_{\bf k})$ is an on-shell four-vector.

It is convenient  to express the initial state in the
 $P$ representation
\begin{eqnarray}
\hat{\rho} = \int Dz f_P(z) |z\rangle \langle z|, \label{psymbol}
\end{eqnarray}
 where the functional $f_P(z)$ is   the $P$-symbol of  quantum state $\hat{\rho}$ \cite{Kla85}. The functional integration in Eq. (\ref{psymbol}) is defined in terms of the Gaussian  measure associated to the Bargmann representation of the quantum field \cite{berezin}.

For simplicity, we assume that no  magnitude $q$ is recorded.
The   probability density Eq. (\ref{prob1aa}) is expressed as
\begin{eqnarray}
P(x) = \int Dz f_P(z) P_z( x),
\end{eqnarray}
where $P_z(x) $ is the probability density Eq. (\ref{prob1aa}) evaluated for an initial  coherent state. It is straightforward to show that  $P_z$ is a sum of three contributions
\begin{eqnarray}
P_z(x) = P_z^{(0)}(x) + P_z^{(1)}( X) + P_z^{(2)}(x), \label{probphoto}
\end{eqnarray}
where
\begin{eqnarray}
P_z^{(0)}(x) &=& \int \frac{d^3k}{(2\pi)^3 } \frac{\tilde{R}(k)}{2\omega_{\bf k}},
\nonumber \\
P_z^{(1)}(x) &=&   \int \frac{d^3k}{(2\pi)^3\sqrt{2\omega_{\bf k}} }   \frac{d^3k'}{(2\pi)^3\sqrt{2\omega_{\bf k}'} } z(k) z^*(k') e^{i(k-k')\cdot x} \tilde{R}\left(\frac{k+k'}{2}\right),
\nonumber \\
P_z^{(2)}( x) &=& 2 \mbox{Re} \; \int \frac{d^3k}{(2\pi)^3\sqrt{2\omega_{\bf k}} }   \frac{d^3k'}{(2\pi)^3\sqrt{2\omega_{\bf k}'} } z(k) z(k')e^{i(k+k')\cdot x}\tilde{R}\left(\frac{k' - k}{2}\right),
\end{eqnarray}
The  three terms in
  Eq. (\ref{probphoto}) have the following interpretation.

  \begin{itemize}

\item   The term $P_z^{(0)}$ is  essentially vacuum noise. It is constant and state independent, and it contributes negligibly to the total detection probability.

  \item The  term $P_z^{(1)}$ is generated by the components of the coherent states that is diagonal with respect to the total particle number.

  \item The  term  $P_z^{(2)}$ involves contributions from terms with different numbers of photons.

  \end{itemize}

The term  $P_z^{(1)}$ in  Eq. (\ref{probphoto}) involves a double momentum integral over $e^{-i(\omega_k - \omega_{k'})t}$ (co-rotating terms), while the   term  $P_z^{(2)}$ involves a double integral over $e^{-i(\omega_k + \omega_{k'})t}$ (counter-rotating terms). For sufficiently large $t$, co-rotating terms typically dominate over counter-rotating terms, hence, we can invoke the  Rotating Wave Approximation (RWA) and drop the contribution of the latter terms.

The RWA is a defining approximation of Glauber's photo-detection theory. Its domain of validity   in photodetection has been  a matter of some debate. Photodetection probabilities obtained from the RWA appear to violate
 causality at short times  \cite{ByTa, Fleisch}. However, it has been suggested that a modified form of the RWA in photodetection   can guarantee causality \cite{MJF95}.  Fundamentally, RWA is unacceptable because the corresponding Hamiltonian  is unbounded from below  \cite{FoCo}. For   details about the accuracy of the RWA and related approximations, see, Ref. \cite{FCAH10} and references therein.

To examine the validity of the RWA in the QTP probability assignment, we consider a coherent state that describes a pulse of mean wave-number ${\bf k}_0$ initially ($t = 0$) localized around ${\bf x} = 0$.   We choose a Gaussian
\begin{eqnarray}
z(k) = (2\pi)^3  z_0 (2 \pi \Delta^2)^{-3/2} e^{ - \frac{ ({\bf k} - {\bf k}_0)^2}{2 \Delta^2}},
\end{eqnarray}
where $z_0$ is a constant complex valued vector and $\Delta$ is the spectral width of the pulse.

First, we evaluate the term $P_z^{(1)}(\tau, {\bf Q})$ in Eq. (\ref{probphoto}) using  a saddle point approximation. We expand $\omega_k = \omega_{k_0} + {\bf v}_{k_0} \cdot ({\bf k} - {\bf k}_0)$, where  ${\bf v}_k = {\bf k}/|{\bf k}|$ is the three- velocity vector. We also assume  that $\tilde{R}$ varies slowly with $k$, so that the contribution of the detector kernel is a constant $\tilde{R}(k_0)$. We obtain
\begin{eqnarray}
P_z^{(1)}(t, {\bf x}) = \frac{1}{2} |z_0|^2 \tilde{R}(k_0) e^{- \Delta^2 ({\bf x} - {\bf v}_{k_0} t)^2}.
\end{eqnarray}
We also evaluate the term $P_z^{(2)} $ in the saddle-point approximation. We find
\begin{eqnarray}
P_z^{(2)}(t, {\bf x}) \sim \mbox{Re} \left[ z_0^2 e^{i k_0\cdot x }\right]. \label{p2rwa}
\end{eqnarray}
The oscillatory terms in Eq. (\ref{p2rwa}) must be averaged over a spatial region of size $\delta_x^3$ and for times of order $\delta_t$. For Gaussian smearing functions, this averaging leads to  multiplicative  factors of order $e^{-\delta_t^2 |{\bf k}|_0^2} e^{-\delta_x^2 \omega_{k_0}^2}$. Hence, for sufficiently large ${\bf k}_0$, the contribution of the term $P_z^{(2)}$ to the probability density (\ref{probphoto}) is suppressed, and the RWA  is justified. In general, we expect the RWA fails only for states with significant contribution from photons in the deep infrared.

 Assuming the RWA, the  detection probability  can be expressed as
\begin{eqnarray}
P(x) = \int d^4 y R(y)  G_{RWA}[x - \frac{1}{2}y, x+\frac{1}{2}y], \label{pqtem2}
\end{eqnarray}
in terms of the correlation function
\begin{eqnarray}
G_{RWA}(x, x') =  2 \langle \Psi|\hat{\phi}^{(-)}(x') \hat{\phi}^{(+)}(x) |\Psi\rangle,
\end{eqnarray}
where $\hat{\phi}^{(+)}$ and $\hat{\phi}^{(-)}$ denote the positive and negative frequency part of $\hat{\phi}$.

Suppose now that the spacetime spread of $R(x)$ is much smaller than the spatial and temporal variation of the field. For massless particles, this is a stronger form of the dipole approximation that is often employed in quantum optics: we assume that the typical wavelengths and frequencies in the initial state are much larger than the size of the detector and the associated time spread of detection. Then, we can approximate $R(x)$ with a delta function, to obtain
\begin{eqnarray}
P( x) \sim  G_{RWA}[x, x].
\end{eqnarray}
This expression coincided with the  standard formula of Glauber's photo-detection theory.

Hence, QTP  recovers Glauber's probabilities for a single measurement event by assuming (i) the RWA   and (ii) $R(y) \sim \delta (y)$. As we showed, for a large class of initial states, the RWA is not an independent assumption,
 as it arises due to the spatial and temporal coarse-graining of the detector. Thus, the crucial assumption of Glauber's theory is that the characteristic scales of the detector kernel  are much  smaller than any time / length parameter that characterize the initial state of the field.
A more detailed analysis, with an emphasis on the divergence from Glauber's theory, will be undertaken elsewhere.

\section{Links to non-equilibrium QFT}

In Sec. 2, we saw that the QTP probabilities are linear functional of balanced correlation functions. The measurements do not probe unbalanced correlation functions. Since the latter include $\langle \hat{C}_a(x)\rangle$, QTP probabilities cannot access mean field information. This limitation is not fundamental. Remember that the operator $\hat{C}(x)$ appears in the interaction term with the apparatus. This restriction means that we cannot use couplings of the form $\int d^4x \hat{C}_a(x) \otimes \hat{J}^a(X)$, in order to directly measure the operator $\hat{C}_a(x)$. At least such measurements are not possible with weak field-apparatus coupling where perturbation theory is applicable\footnote{It is possible to measure $\hat{C}_a(x)$ for non-perturbative couplings \cite{PapAna22}, but the resulting probabilities are very different from those of QTP, and they are closer to those of von Neumann measurements.}.

Suppose, for example that $\hat{C}$ coincides with the field operator $\hat{\phi}$---we drop the index $a$ for simplicity.  Then, single-detector probabilities record only local information about particles. Let the field be  in a state characterized by  a macroscopically large number of particles; then, it can be viewed as a thermodynamic system. Then the single-detector probability essentially coincides with a particle-number density function. If we also measure the recorded particle's momentum $k$, the QTP probability density $P(x, k)$ is an operationally defined version of Boltzmann's distribution function.
By Eq. (\ref{prob1aa}), $P(x, k)$ is a linear functional of the correlation function $G(x, x') = \langle \hat{\phi}(x) \hat{\phi}(x')\rangle$, which is usually taken to satisfy the Baym-Kadanoff equations.

From the above analysis, it follows that  Boltzmann's thermodynamic entropy, defined on a Cauchy surface $\Sigma$,
\bey
S_{B}(\Sigma) = -  \int_{\Sigma} d^3 x d^3k P(x,k) \ln P(x,k) \label{boltzent}
\eey
 is a Shannon-type entropy for  single-detection measurements. This means that quantum informational quantities, defined through measurements, have a direct application to non-equilibrium QFT. Furthermore, $n$-detector QTP probabilities probe higher-order correlation functions of the quantum field, thus allowing an analysis that is not accessible by traditional methods.

Such an analysis is beyond the scope of this paper. In this section, we will only describe some structural similarities and connections between the QTP analysis and methods of non-equilibrium QFT. A detailed analysis will be taken up in other publications.

\subsection{Stochastic correlation dynamics from two-particle irreducible effective action}
As shown in Sec. 3.2, the generating functional of QTP correlation functions is defined in terms of non-local source terms $L^{ab}(x, x')$. It is structurally similar to the two-particle irreducible  effective action (2PIEA) \cite{ctp5,RH97} that has found many applications in non-equilibrium QFT---see, for example, \cite{CH08, Berges, Berges2}.

We proceed to a brief  review of the 2PIEA formalism following \cite{cddn, CH99}, emphasizing how it can be used to define evolution equations with noise from higher-order correlation functions.

For ease of notation, we use a version of DeWitt's condensed notation, where capital indices $A$ correspond to both the spacetime dependence and the branch of the CTP field (forward or backward in time, $\phi$ or $\bar{\phi}$). Hence, we will be writing $\phi_A$, $C_A$, and so on. The action in the CTP generating functional will be $S[\phi_A] = S[\phi_a] - S[\bar{\phi}_a]$. In the two-particle irreducible representation,   the (two-point) correlation function stands is an independent variable, not a functional of the mean field. Thus there is a  separate source $K^{AB}$ driving $C_A C_B$ over the usual $J^A C_A$ term in the one-particle irreducible representation---see the similarity to Eq. (\ref{ZQTP}).

From the generating functional
\begin{equation}
Z\left[ K^{AB}\right] =e^{iW\left[ K^{AB}\right] }=\int D\phi_A
\;e^{i\left( S+\frac 12K^{AB}C_AC_B\right) }  \label{genfun}
\end{equation}
we have
\begin{equation}
G^{AB}=\left\langle \hat{C}_A\hat{C}_B\right\rangle =2\left. \frac{\delta W}{%
\delta K^{AB}}\right| _{K=0}  \label{greenfun}
\end{equation}
and
\begin{equation}
\left. \frac{\delta^2W}{\delta K^{AB}\delta K^{CD}}\right| _{K=0}=\frac{i}{4}%
\left\{ \left\langle \hat{C}_A \hat{C}_B \hat{C}_C \hat{C}_D \right\rangle -\left\langle
 \hat{C}_A \hat{C}_B \right\rangle \left\langle\hat{C}_C \hat{C}_D\right\rangle \right\}.
\label{fluc}
\end{equation}
Suppose that we want to express the effective dynamics of $G_{AB}$ in a closed form, but to go beyond the Baym-Kadanoff equations, by taking into account noise from higher-order correlations.
For an non-equilibrium system, we seek a  formulation in terms of  a new object  ${\bf G}_{AB}$. This is  a stochastic correlation function whose expectation value over the noise average  gives the usual two point functions. The fluctuations of ${\bf G}_{AB}$ reproduce the quantum fluctuations in the binary products of field operators. The simplest assumption is to take  ${\bf G}_{AB}$ as a Gaussian process, defined
by

\begin{equation}
\left\langle {\bf G}_{AB}\right\rangle =\left\langle \hat{C}_A\hat{C}_B\right\rangle ;\qquad \left\langle {\bf G}_{AB}{\bf G}_{CD}\right\rangle
=\left\langle \hat{C}_A \hat{C}_B \hat{C}_C \hat{C}_D\right\rangle  \label{stocg}
\end{equation}


The Legendre transform of $W$ is the two-particle irreducible effective action,
\begin{equation}
\Gamma_{2PI} \left[ G_{AB}\right] =W\left[ K^{AB}\right] -\frac 12K^{AB}%
{G}_{AB};\qquad K^{AB}=-2\frac{\delta \Gamma }{\delta {G}_{AB}}
\label{tpiea}
\end{equation}
The Schwinger-Dyson equation for the propagators is simply, $\frac{\delta \Gamma_{2PI} }{\delta G_{AB}}=0$. When including the stochastic source  ${\bf G}_{AB}$, it becomes
\begin{equation}
\frac{\delta \Gamma_{2PI} }{\delta {\bf G}_{AB}}=-\frac{1}2\kappa^{AB}
\label{lan2pi}
\end{equation}
where $\kappa _{ab}$ is a stochastic nonlocal Gaussian source defined by
\begin{equation}
\left\langle \kappa^{AB}\right\rangle =0;\qquad \left\langle \kappa^{AB}\kappa^{CD}\right\rangle =4i\left[ \frac{\delta ^2\Gamma_{2PI} }{\delta
G_{AB}\delta G_{CD}}\right] ^{\dagger }  \label{noisecor}
\end{equation}

The noiseless Eq. (\ref{lan2pi}) ($\kappa = 0$) provides the
basis for the derivation of transport equations in the near equilibrium
limit. Indeed, for a $\lambda \phi ^4$ theory, we obtain the Boltzmann equation for a distribution function $f$ defined from the Wigner transform of $G^{ab}$.
The full stochastic equation (\ref{lan2pi}) leads, in the same limit, to a Boltzmann - Langevin equation \cite{CH99}.

\subsection{Correlation Histories}
The two-particle irreducible formalism can be extended to an n-particle irreducible formalism, for any $n$. There is an effective action $\Gamma_{nPI}$  for each $n$, from which all effective actions for $n' < n$ can be derived. Taking $n\rightarrow \infty$, we obtain a master effective action. The functional variation of the master effective action  yields the hierarchy of Schwinger-Dyson equations \cite{cddn}.

To obtain effective closed dynamics for the correlations at order $n$, we must truncate the Schwinger-Dyson hierarchy upon this order. Truncation renders the master effective action complex. Its imaginary part arises from correlation functions of order higher than $n$, the fluctuations of which Calzetta and Hu define as  {\em correlation noises} \cite{CH99} at order $n$. For example, the noise $\kappa^{AB}$ in Eq. (\ref{lan2pi}) is the correlation noise of order two.

Calzetta and Hu defined the notion of {\em correlation histories} \cite{DCH}, in analogy to the decoherent histories program. A fine-grained correlation  history corresponds to the full Schwinger-Dyson hierarchy of correlation functions. When we truncate the hierarchy at finite order $n$, we treat only correlation functions of order $n$ as independent. Higher -order correlations are ignored or slaved to  the lowest-order ones.
A truncated hierarchy defines a {\em coarse-grained} correlation history. For example, mean field theory studies {\em coarse-grained} correlation histories at order $n = 1$; the Baym-Kadanoff equation, or Boltzmann equation and their stochastic generalizations  refer to {\em coarse-grained} correlation histories of order $n = 2$.

The key point is that the truncation of the master effective action always leads to dissipation and noise for the coarse-grained histories. Any truncated theory is an effective field theory in the correlation hierarchy formulation. This effective field theory does not carry the full information, this loss of information being expressed as   correlation noise.
The higher-order correlations are analogous to an environment in the theory of open quantum systems \cite{Ana97}. This noise may lead to decoherence of correlation histories \cite{DCH}, i.e., to the classicalization of the effective description.

The QTP approach demonstrates that the different levels of correlation histories can be accessed by the measurement of $n$-detector joint probabilities.
Eq. (\ref{probdenr}) assigns to each initial state $|\psi\rangle$ of the field a hierarchy of joint  probability distributions  $P_n(z_1, z_2, \ldots, z_n)$.

 In classical probability theory, a hierarchy of correlation functions defines a classical stochastic process, if it satisfies the Kolmogorov additivity condition,
\bey
P_{n-1}(z_1, \ldots, z_{n-1}) = \int dz_n  P_n(z_1, z_2, \ldots, z_n) \label{Kolmadd}
\eey
Quantum probability distributions for sequential measurements do not satisfy this condition \cite{Ana06}. Hence, the violation of Eq. (\ref{Kolmadd}) is a genuine signature  of quantum dynamics; it cannot be reproduced by classical physics, including classical stochastic processes. It is rather different from the Leggett-Garg inequalities \cite{LeGa} that also refer to the behavior of quantum multi-time probabilities. Some authors refer to the violation of Eq. (\ref{Kolmadd}) as "temporal non-locality" \cite{Brunnt}.
In contrast, if measurements on a quantum field approximately satisfy Eq. (\ref{Kolmadd}), then the measurement outcomes can be  simulated by a stochastic process with $n$-time probabilities given by the probability distributions (\ref{Kolmadd}). Then the generating functional $Z_{QTP}$ corresponds to a stochastic process, i.e., it is obtained as the functional Laplace transform of a classical stochastic probability measure.

Hence, the hierarchy $P_n(z_1, z_2, \ldots, z_n)$ provides a natural and unambiguous classicality criterion\footnote{For a different classicality criterion, based on the decoherent histories approach, that applies  to the special case of measurements of the field operator, see \cite{Ana03}.}. Given the relation between QTP probabilities and QFT correlation functions, this criterion can be used to probe the information content of different levels for correlation histories. For example, the validity of Eq. (\ref{Kolmadd}) is necessary for deriving deterministic or classical stochastic dynamics for $P_1(z)$, i.e., for deriving the Boltzmann or the Boltzmann-Langevin equation. The failure of (\ref{Kolmadd}) means that the four-point correlation functions are too `quantum' to allow effective  classical  stochastic dynamics for the two-point correlation function.

Conversely, the failure of Eq. (\ref{Kolmadd}) can be used to provide a measure of irreducibly quantum information at the level $n = 2$. An example of such a measure is
\bey
S_Q = \int dz_2 \left|\int dz_1 P_2(z_1, z_2) - P_1(z_2)\right|.
\eey
A second informational quantity is the correlation of the probability distribution, i.e., a measure of the deviation  of $P_2(z_1, z_2)$ from $P_1(z_1) P_2(z_2)$\footnote{The factorization assumption of the joint probability for two events is equivalent to Boltzmann's Stosszahlansatz in the derivation of Boltzmann's equation through the Bogoliubov–Born–Green–Kirkwood–Yvon  hierarchy \cite{Balescu}. The joint probability density $P_2$ is then slaved to $P_1$}. This information is typically quantified by  the correlation entropy
\bey
S_C = \int dz_1 dz_2 P_2(z_1, z_2) \ln \frac{P_2(z_1, z_2)}{P_1(z_1) P_1(z_2)}.
\eey
In general, the correlation entropy will contain information for both quantum correlations (if $S_Q \neq 0$) and classical stochastic ones. Indeed, $S_C$ may not have the usual properties of correlation entropy if $S_Q \neq 0$, and other measures that will distinguish will be more convenient. The third relevant informational quantity in Boltzmann's entropy (\ref{boltzent}), defined in terms of $P_1(z_1)$. These three quantities are the most important for describing the flow of information at the level of the 2PIEA.

 Hence, the QTP hierarchy functions as a registrar of information of the quantum system,   keeping track of how much information resides in what order, and how it flows from one order to another through the dynamics. There is a good potential for this scheme to systemize quantum information in QFT\footnote{Of course, several issues remain to be addressed. The probability distributions $P_n$ depend on the measurement apparatus through the detector kernels. We would like to avoid strong apparatus dependence when deriving informational quantities. We may have to consider ideal, maximal information, apparatuses (like the ones appearing in the study of the time-of-arrival), or look for quantities that can be minimized with respect to all physical detector kernels. }:    keeping track of the contents and the flow of information and   measuring the degree of coherence in a quantum system.

\section{Conclusions}
In this work, we presented the QTP formalism for measurements in quantum fields, and we explored its  relations to the Closed-Time-Path description of QFT. These relations provide  a direct translation between the operational language of measurement theory (POVMs, effects, and so on) to the manifestly covariant description of QFT through functional methods. This is an essential step towards formulating a general theory of relativistic quantum information.

Our work highlights the central role of the hierarchy of unequal-time correlation functions. On one hand, POVMs for measurement are linear functional of such correlation functions, on the other hand, these correlation functions are crucial for the implementation of perturbation theory, but also give real causal dynamics  in the CTP approach. When working at the level of non-equilibrium QFT, correlation functions are employed in order to define both observables and effective irreversible dynamics.

Equally important is the fact that correlation functions are covariant and causal objects. For this reason, we contend that a relativistic QIT that respects both causality and spacetime symmetry must define all informational quantities in terms of such correlation functions. This contrasts the standard approach of QIT
 that is based on properties of single-time quantum states, like von Neumann entropy or entanglement.

 Building a sound theoretical foundation for relativistic QIT is not only important for theoretical completeness. It is also needed for describing quantum experiments in space \cite{Rideout, DSQL2} that will explore the effects of   non-inertial motion (acceleration, rotation) and gravity on quantum resources like entanglement.  Corrections to standard photodetection models, like the ones predicted by QTP, may be measurable  in experiments that involve long separations or large relative velocities between detectors.

\section*{Acknowledgements}
Research was supported by   grant  JSF-19-07-0001 from the Julian Schwinger Foundation.

\begin{appendix}
\section{Derivation of the QTP probability formula}
In this section, we present a brief account of the QTP method, leading to the derivation of Eq. (\ref{probX20}), as a genuine probability density.
\subsection{Amplitudes of detection}
We consider a composite physical system that consists of a microscopic and a macroscopic component. The microscopic component is the quantum system to be measured and the macroscopic component is the measuring device.

 Let ${\cal H}$ be the Hilbert space of the composite system, and $\hat{H}$ the associated Hamiltonian operator. In QTP, we model a measurement event as a
transition between two complementary subspaces of ${\cal H}$. To this end, we split
 ${\cal H}$  in two subspaces: ${\cal H} = {\cal
H}_+ \oplus {\cal H}_-$.
The subspace ${\cal H}_+$ describes the accessible states of the system given that the event under consideration is realized. For example, if the event is a detection of a microscopic particle by  an  apparatus,   ${\cal H}_+$ includes all states of the apparatus with a definite macroscopic record of detection.  We
denote  the projection operator onto ${\cal H}_+$ as $\hat{P}$ and the projector onto ${\cal H}_-$ as $\hat{Q} := 1  - \hat{P}$. We also assume that a pointer variable
 $\lambda$ of the measurement apparatus takes a definite value after the transition has occurred. The corresponding positive operators
$\hat{\Pi}(\lambda)$  satisfy $\hat{\Pi}(\lambda)$ satisfy
 $  \sum_\lambda \hat{\Pi}(\lambda) = \hat{P}$.

 First, we construct
 the probability amplitude $| \psi; \lambda, [t_1, t_2] \rangle$ that, given a state $|\psi_0\rangle \in {\cal H}_-$ at $t = 0$, a transition occurs during the time interval $[t_1, t_2]$ and a value $\lambda$ is obtained for the pointer variable.
 For a vanishingly small time
interval, we set $t_1 = t$ and $t_2 = t + \delta t$, and we  keep only leading-order terms with respect to $\delta t$. At times prior to $t$, the state
lies in ${\cal H}_-$. This is taken into account by evolving the initial state $|\psi_0 \rangle$ with the restricted propagator in ${\cal H}_-$,
 \begin{eqnarray}
 \hat{S}_t =  \lim_{N
\rightarrow \infty} (\hat{Q}e^{-i\hat{H} t/N} \hat{Q})^N. \label{restricted}
\end{eqnarray}
By assumption, the transition occurs at some instant within the time interval $[t, t+\delta t]$, hence, there is no constraint in the propagation from $t$ to $t + \delta t$; the propagation  is implemented by  the unrestricted evolution operator   $e^{-i \hat{H} \delta t} \simeq 1 - i \delta t \hat{H}$. At time $t + \delta t$, the event corresponding to $\hat{\Pi}_{\lambda}$ is recorded, so the
amplitude is transformed by the action of $\hat{\Pi}_{\lambda}$ ($\sqrt{\hat{\Pi}_{\lambda}}$, if $\hat{\Pi}_{\lambda}$ is not a projector). For times greater than $t + \delta t$, there is no constraint, so the amplitude
evolves as $e^{-i \hat{H} (T-t)}$  until some final moment $T$.

At the limit of small $\delta t$, the successive operations above yield
\begin{eqnarray} |\psi_0; \lambda, [t, t+ \delta t] \rangle =  - i \, \delta t \, \,e^{-i\hat{H}T }  \hat{C}(\lambda, t) |\psi_0
\rangle. \label{amp1} \end{eqnarray}
where
\bey
 \hat{C}(\lambda, t) := e^{i \hat{H}t} \sqrt{\hat{\Pi}}(\lambda)\hat{H}\hat{S}_t.
\eey
is a history operator.   Since  the amplitude (\ref{amp1}) is proportional to $\delta t$, it defines  a {\em density} with respect to time: $|\psi_0;  \lambda, t \rangle := \lim_{\delta t \rightarrow 0}
\frac{1}{\delta t} | \psi_0; \lambda, [t, t + \delta t] \rangle$. From Eq. (\ref{amp1})

The total amplitude that the transition occurred at {\em some moment} within a time interval $[t_1, t_2]$ is
\begin{eqnarray}
| \psi; \lambda, [t_1, t_2] \rangle = - i e^{- i \hat{H}T} \int_{t_1}^{t_2} d t \hat{C}(\lambda, t) |\psi_0 \rangle. \label{ampl5}
\end{eqnarray}

Eq. (\ref{ampl5}) involves the restricted propagator $\hat{S}_t$. This quantity may be difficult to evaluate in practice. Furthermore, if $\hat{S}_t$ is literally defined by Eq. (\ref{restricted}), it is unitary in
${\cal H}_-$, and may lead to quantum-Zeno-type problems in the definition of probabilities. In Ref. \cite{HalYe} it was shown that these problems can be avoided by employing alternative
 definitions of the restricted propagator, that involve a regularization time scale. However, this renders $\hat{S}_t$ model dependent.

 However, for the type of measurements considered here, the problems with the evaluation of
$\hat{S}_t$ can be avoided.    We consider a Hamiltonian   $\hat{H} = \hat{H_0} + \hat{H_I}$ where
$[\hat{H}_0, \hat{P}] = 0$, and $H_I$ is a perturbing interaction. To leading order in the interaction,
\begin{eqnarray}
 \hat{C}(\lambda, t) = e^{i \hat{H}_0t} \sqrt{\hat{\Pi}}(\lambda) \hat{H}_I e^{-i \hat{H}_0t}, \label{classpert}
\label{perturbed}
\end{eqnarray}
with no dependence on $\hat{S}_t$.

 \subsection{Probabilities of detection}

 We define the probability $\mbox{Prob} (\lambda, [t_1, t_2])\/$ that at some time in the interval $[t_1, t_2]$ a detection with outcome $\lambda$ occurred by taking the squared modulus of the amplitude  Eq. (\ref{ampl5}),
\begin{eqnarray}
\mbox{Prob}(\lambda, [t_1, t_2])
:= \langle \psi; \lambda, [t_1, t_2] | \psi; \lambda, [t_1, t_2] \rangle =   \int_{t_1}^{t_2} \,  dt \, \int_{t_1}^{t_2} dt' Tr\left[\hat{C}(\lambda, t)
\hat{\rho_0}\hat{C}^{\dagger}(\lambda, t')\right], \label{prob1}
\end{eqnarray}
where $\hat{\rho}_0 = |\psi_0\rangle \langle \psi_0|$.

The quantities $\mbox{Prob} (\lambda, [t_1, t_2])\/$ do not define a probability measure with respect to time, because they do not satisfy the Kolmogorov additivity condition $\mbox{Prob}(\lambda, [t_1, t_3]) = \mbox{Prob}(\lambda, [t_1, t_2]) + \mbox{Prob}(\lambda, [t_2, t_3])$ fails, unless the quantity
\begin{eqnarray}
{\cal D}(t_1, t_2, t_3):=  2 Re \left[ \int_{t_1}^{t_2} \,  dt \, \int_{t_2}^{t_3} dt' Tr\left(\hat{C}(\lambda, t) \hat{\rho_0}\hat{C}^{\dagger}(\lambda, t')\right)\right] \label{decond}
 \end{eqnarray}
 vanishes. Clearly, ${\cal D}(t_1, t_2, t_3)$ does not vanish identically.   However, in a macroscopic system (or in a system with a macroscopic component) one expects that Eq. (\ref{decond}) holds with a good degree of approximation, given a sufficient degree of coarse-graining \cite{GeHa2, hartlelo}---see also the explicit measurement models in \cite{QTP1}.
   Thus, a necessary condition for the time of transition to be associated to macroscopic records in a measurement apparatus is the existence of a
    coarse-graining time-scale $\sigma$, such that  $|{\cal D}(t_1, t_2, t_3)|$ is negligible
    for  $ |t_2 - t_1| >> \sigma$ and $|t_3 - t_2| >> \sigma$. Then, Eq. (\ref{prob1}) does define a probability measure when restricted to intervals of size  larger than $\sigma$.

We can express the time-of-transition  probabilities in terms of
 densities
 with respect to a continuous time variable by smearing
 the amplitudes
 Eq. (\ref{ampl5}) with respect to a time-scale  $\delta_t >> \sigma$.
To this end, we introduce a family of functions $f(s)$ of width $\delta_t$,  localized around $s = 0$, and normalized so that $\lim_{\delta_t \rightarrow 0} f(s) = \delta(s)$.
Then, we  define the smeared amplitude
\begin{eqnarray}
|\psi_0; \lambda, t\rangle_{\delta_t} := \int ds \sqrt{f(s -t)} |\psi_0; \lambda, s \rangle =  \int ds \sqrt{f(s - t)} \hat{C}(\lambda, s) |\psi_0 \rangle. \label{smearing}
\end{eqnarray}

 The modulus-
squared amplitudes
\begin{eqnarray}
 W(t, \lambda) = {}_{\sigma}\langle \psi_0; \lambda, t|\psi_0; \lambda, t\rangle_{\sigma} = \int ds ds' \sqrt{f(s-t) f(s'-t)} Tr \left[\hat{C}(\lambda, s)
 \hat{\rho}_0 \hat{C}^{\dagger}(\lambda, s')\right] \label{ampl6}
\end{eqnarray}
 define a probability measure: they are of the form
 $Tr[\hat{\rho}_0 \hat{\Pi}(\lambda, t)]$,
where
\begin{eqnarray}
\hat{\Pi}(\lambda, t) = \int ds ds' \sqrt{f_{\sigma}(s-t) f_{\sigma}(s'-t)} \hat{C}^{\dagger}(\lambda, s') \hat{C}(\lambda, s) \label{povm2}
\end{eqnarray}
is a density with respect to both variables
$\lambda$ and $t$.

The positive operator
\begin{eqnarray} \hat{\Pi}_{\tau}(N) = 1 - \int_0^{\infty} dt \int d \lambda \hat{\Pi}_{\tau}(\lambda, t), \label{nodet}
\end{eqnarray}
 corresponds to the alternative $ N$ that no detection took place
in the time interval $[0, \infty)$. Since the detector is a quantum system, there exists a non-zero probability that the microscopic particle excites no transition, and thus, no record is left.

The operator $\hat{\Pi}_{\tau}(N)$ together with the positive operators Eq. (\ref{povm2})
 define a Positive-Operator-Valued Measure (POVM). The POVM Eq. (\ref{povm2}) determines the probability density that a transition took place at time $t$, and that the outcome
$\lambda$ for the value of an observable has been recorded.

Consider a Hilbert space  of the form ${\cal F} \otimes {\cal K}$, where ${\cal F}$ describes a quantum field and ${\cal K}$ describes the apparatus. We assume the same set-up as in Sec. 2.3, except for the fact that the interaction Hamiltonian is Poincar\'e invariant,  i.e., it does not involve a switching function,
$\hat{H}_I = \int d^3x \hat{C}_a({\bf x}) \hat{J}^a({\bf x})$. Then,
the probability density (\ref{ampl6}) contains a kernel $\langle \Omega|\hat{J}^a({\bf x},t) \hat{\Pi}(\lambda) \hat{J}^b({\bf x}',t') |\Omega\rangle$. We take the pointer variable $\lambda$ to describe the position ${\bf X}$ of the detection record and another observable $q$. Since $\hat{\Pi}(\lambda) $ is a highly coarse-grained observable, we can approximate it as a product $\hat{\Pi}({\bf X}) \hat{\Pi}(\lambda)$, i.e., a product of two commuting POVMs,  for position and for the observable $q$. Then, we can write $\hat{\Pi}(\lambda) = \sqrt{\hat{\Pi}({\bf X})} \hat{\Pi}(\lambda)\sqrt{\hat{\Pi}({\bf X})}$.

Suppose we sample position with accuracy $\delta_x$. Then we expect that $\sqrt{\hat{\Pi}({\bf X})} \hat{J}^a({\bf x},t) |\Omega\rangle$ vanishes if $|{\bf X} -{\bf x}| >> \delta_x$. Hence, we can write $\sqrt{\hat{\Pi}({\bf X})} \hat{J}^a({\bf x},t) |\Omega\rangle = C \sqrt{g}({\bf X} -{\bf x}) \hat{J}^a({\bf x},t) |\Omega\rangle$, where $g$ is a positive sampling function of width $\delta_x$ and $C$ is a multiplicative constant. With this substitution, we recover Eq. (\ref{probX20}), only now rather than the switching function $F(x - x')$, we have the sampling functions $\sqrt{f(t - t') g({\bf x} - {\bf x'})}$.

\end{appendix}


\begin{thebibliography}{}

\small

\bibitem{AnSav22} C. Anastopoulos and N. Savvidou,	{\em Quantum Information in Relativity: The Challenge of QFT Measurements},  Entropy 24, 4 (2022).

\bibitem{Rideout}	D. Rideout et al, {\em Fundamental Quantum Optics Experiments Conceivable with Satellites -- Reaching Relativistic Distances and Velocities}, Class. Quantum Grav. 29, 224011 (2012).


\bibitem{DSQL2}M. Mohageg et al, {\em The Deep Space Quantum Link: Prospective Fundamental
Physics Experiments using Long-Baseline Quantum Optics}, arXiv:2111.15591.

\bibitem{AnHu15} C. Anastopoulos and B. L. Hu, {\em Probing a Gravitational Cat State}, Class. Quant. Grav. 32, 165022 (2015).

\bibitem{Bose17} S. Bose, A. Mazumdar, G. W. Morley, H. Ulbricht, M. Toro, M. Paternostro, A. A. Geraci, P. F. Barker, M. S. Kim, and G. Milburn, {\em A Spin Entanglement Witness for Quantum Gravity}, Phys. Rev. Lett. 119, 240401 (2017).

\bibitem{Vedral17} C. Marletto and V. Vedral, {\em Gravitationally Induced Entanglement between Two Massive Particles is Sufficient Evidence of Quantum Effects in Gravity}, Phys. Rev. Lett. 119, 240402 (2017).

\bibitem{AnHu20} C. Anastopoulos and B. L. Hu, {\em Quantum Superposition of Two Gravitational Cat States}, Class. Quant. Grav. 37, 235012 (2020).




\bibitem{Bloch}	I. Bloch, {\em Some Relativistic Oddities in the Quantum Theory of Observation}, Phys. Rev. 156, 137 (1967).

  \bibitem{Aharonov}   Y. Aharonov and D. Z. Albert, {\em Can we make sense out of the measurement process in relativistic quantum mechanics?}, Phys. Rev. D4, 359 (1981).

\bibitem{PeTe} A. Peres and D. Terno,   {\em Quantum Information and Relativity Theory}, Rev. Mod. Phys. 76, 93 (2004).

\bibitem{WH65} R. M F. Houtappel, H. van Dam,  and E. P. Wigner,  {\em The Conceptual Basis
and Use of the Geometric Invariance Principles}, Rev. Mod. Phys. 37, 595 (1965).



\bibitem{Malament} 	D. Malament, {\em In Defense of Dogma: Why There Cannot Be A Relativistic Quantum Mechanics of (Localizable) Particles}, in "Perspectives on Quantum Reality", ed. R. Clifton (Kluwer Academic, Dordrecht 1996).




  \bibitem{Heg1}  G. C.  Hegerfeldt, {\em Instantaneous Spreading and Einstein Causality in Quantum Theory}, Annalen der Physik 7, 716 (1998).



\bibitem{Sorkin} R. Sorkin, {\em Impossible Measurements on Quantum Fields}, in “Directions in General Relativity”, eds. B. L. Hu and T. A. Jacobson (Cambridge University Press, 1993).


\bibitem{BJK}L. Borsten, I. Jubb, and G. Kells, {\em Impossible measurements revisited},
Phys. Rev. D104, 025012 (2021).

\bibitem{Peres2} A. Peres,    {\em Classical interventions in quantum systems. II. Relativistic invariance}, Phys. Rev. A61,  022117 (2000). .

\bibitem{PoVai}S. Popescu, and L. Vaidman, {\em   Causality constraints on nonlocal quantum measurements},
Phys. Rev. A49,   4331 (1994).

\bibitem{InstMeas} L. Vaidman,  {\em  Instantaneous Measurement of Nonlocal Variables},
Phys. Rev. Lett. 90,  010402 (2003).




\bibitem{GoPr} D. Beckman, D.  Gottesman, M. A. Nielsen, J. Preskill,   { \em Causal and localizable quantum operations},  Phys. Rev. A64,    052309 (2001).

\bibitem{Grimmer} D. Grimmer, {\em The Pragmatic QFT Measurement Problem and the need for a Heisenberg-like Cut in QFT },  	arXiv:2205.09608.


\bibitem{LP31} L. Landau and R. Peierls, {\em Erweiterung des Unbestimmtheitsprinzips für die Relativistische Quantentheorie},  Zeit. Phys. 69, 56 (1931).

\bibitem{BoRo} N. Bohr and L. Rosenfeld, {\em On the Question of the Measurability of Electromagnetic Field Quantitie}, Mat.-fys. Medd. Dan. Vid. Selsk.12, (1933).


\bibitem{Glauber1} 	R. J. Glauber, {\em The Quantum Theory of Optical Coherence}, Phys. Rev. 130, 2529 (1963).

 \bibitem{Glauber2} 	R. J. Glauber,{\em Coherent and Incoherent States of the Radiation Field}, Phys. Rev. 131, 2766 (1963).


\bibitem{QuOp} M. O. Scully and M. S. Zubairy, {\em Quantum Optics} (Cambridge University Press, 2012).



\bibitem{RWA1} G. S. Agarwal, {\em Rotating-Wave Approximation and Spontaneous Emission}, Phys. Rev. A4, 1778 (1971);
  Phys. Rev. A7, 1195 (1973)

\bibitem{RWA2}  C. H. Fleming, N. I. Cummings, C. Anastopoulos and B. L. Hu, {\em The Rotating-Wave Approximation: Consistency and Applicability from an Open Quantum System
Analysis}, J. Phys. A: Math. Theor.
43, 405304 (2010).


\bibitem{phc1}M. de Haan, {\em Photodetection and causality II}, Physica 132A, 375, 397 (1985);  V.P.Bykov and V.I.Tatarskii, {\em Causality Violation in the Glauber Theory of Photodetection}, Phys.Lett.A136, 77 (1989); V.I.Tatarskii, {\em Corrections to the theory of photocounting}, Phys.Lett. A144, 491 (1990).

\bibitem{phc2} L. I. Plimak and S. Stenholm, {\em Causal signal transmission by quantum fields. II: Quantum-statistical response of interacting bosons}, Ann.Phys. 323, 1989 (2008); L. I. Plimak, S. T. Stenholm, and W. P. Schleich, {\em Operator ordering and causality}, Phys. Scr. 2012, 014026  (2012).

\bibitem{FuMa} N. Funai and E. Martin-Martinez, {\em  Faster-than-light signaling in the rotating-wave approximation}, Phys. Rev. D100, 065021 (2019).

\bibitem{Unruh76} W. G. Unruh, {\em Notes on Black Hole Evaporation}, Phys. Rev. D14, 870 (1976).

\bibitem{Dewitt}B. S. DeWitt, {\em Quantum Gravity: the New Synthesis} in
General Relativity: An Einstein Centenary Survey, ed. by S. W. Hawking
and W. Israel (Cambridge University Press, Cambridge, 1979), p.
680.


\bibitem{LoSa}  J. Louko and A. Satz, {\em
How often does the Unruh–DeWitt detector click? Regularization by a spatial profile }, Class. Quantum Grav. 23, 6321 (2004).

\bibitem{Perche} T. R. Perche, {\em Localized nonrelativistic quantum systems in curved spacetimes: A general characterization of particle detector models},
Phys. Rev. D106, 025018 (2022).


\bibitem{GGM22}  J. Polo-Gómez, L. J. Garay, L. J. and E. Martín-Martínez,  {\em A detector-based measurement theory for quantum field theory},  	Phys. Rev. D 105, 065003 (2022).


\bibitem{HLL12} B. L. Hu, S-Y Lin, J. Louko, {\em Relativistic Quantum Information in Detectors–Field Interactions}, Class. Quantum Grav. 29, 224005 (2012).


\bibitem{sl1} M. Cliche and A.Kempf, {\em The relativistic quantum channel of communication through field quanta}, Phys. Rev. A 81, 012330 (2010).

 \bibitem{sl2}  E. Martin-Martinez, {\em Causality issues of particle detector models in QFT and Quantum Optics}, Phys. Rev. D 92, 104019 (2015).


\bibitem{sl4}  J. de Ramon, M. Papageorgiou and E. Martin-Martinez, {\em Relativistic causality in particle detector models: Faster-than-light signalling and "Impossible measurements" }, Phys. Rev. D 103, 085002 (2021).


\bibitem{HeKr} K. E. Hellwig and K. Kraus, {\em Formal description of measurements in local quantum field theory}, Phys.
Rev. D 1, 566 (1970).

\bibitem{OkOz} K. Okamura and M. Ozawa, {\em Measurement theory in local quantum physics}, J. Math. Phys. 57, 015209
(2015).

\bibitem{Dop}S. Doplicher, {\em The measurement process in local quantum physics and the EPR paradox}, Commun.
Math. Phys. 357, 407 (2018).


\bibitem{FeVe} C. J. Fewster and R. Verch, {\em Quantum fields and local measurements}, Comm. Math. Phys. 378, 851 (2020).

\bibitem{Few19} C. J. Fewster, {\em A generally covariant measurement scheme for quantum field theory in curved spacetimes}, arXiv:1904.06944.



\bibitem{Ruep} H. Bostelmann, C. J. Fewster and M. H.  Ruep,  {\em Impossible measurements require impossible apparatus}, Phys. Rev. D103, 025017 (2021).

\bibitem{FJR22}C. J. Fewster, I. Jubb, and M. H. Ruep, {\em Asymptotic measurement schemes for every observable of a quantum field theory}, arXiv: 2203.09529.




\bibitem{QTP1}	C. Anastopoulos and N. Savvidou, {\em Time-of-Arrival Probabilities for General Particle Detectors}, Phys. Rev. A86, 012111 (2012).

\bibitem{QTP2}	C. Anastopoulos and N. Savvidou, {\em Time-of-Arrival Correlations}, Phys. Rev. A95, 032105 (2017).

\bibitem{QTP3}	C. Anastopoulos and N. Savvidou, {\em Time of arrival and Localization of Relativistic Particles}, J. Math. Phys. 60, 0323301 (2019).



 \bibitem{AnSav06} C. Anastopoulos and N. Savvidou, {\em Time-of-arrival Probabilities and Quantum Measurements},  J. Math. Phys. 47, 122106 (2006).



\bibitem{AnSav08} C. Anastopoulos and N. Savvidou, {\em Time-of-arrival probabilities and quantum measurements. II. Application to tunneling times}, J. Math. Phys. 49, 022101 (2008).





\bibitem{An08} C. Anastopoulos, {\em Time-of-arrival probabilities and quantum measurements. III. Decay of unstable states}, J. Math. Phys. 49, 022103 (2008).



    \bibitem{Sav99} K. Savvidou,  {\em The Action Operator for Continuous-time Histories}  J. Math. Phys. 40, 5657 (1999); {\em Continuous Time in Consistent Histories}, gr-qc/9912076.

\bibitem{Sav10} N. Savvidou, {\em Space-time Symmetries in  Histories Canonical Gravity}, in "Approaches to Quantum Gravity", edited by D. Oriti (Cambridge University Press, Cambridge 2009).



\bibitem{Gri} R. B. Griffiths, {\em 	Consistent Quantum Theory} (Cambridge University Press, Cambridge 2003).
\bibitem{Omn1} R. Omn\'es, {\em The Interpretation of Quantum Mechanics}, (Princeton University Press, Princeton 1994).

\bibitem{Omn2}  R. Omn\'es,	{\em Understanding Quantum Mechanics} (Princeton University Press, Princeton 1999).




\bibitem{GeHa1} M. Gell-Mann and J.  B.  Hartle,  {\em Quantum Mechanics in
the Light of Quantum Cosmology}, in `Complexity, Entropy, and the Physics of Information',   ed. by W. Zurek,
(Addison Wesley, Reading 1990);

\bibitem{GeHa2} M. Gell-Mann and J.  B.  Hartle, {\em Classical Equations
for Quantum Systems},  Phys. Rev.   D47, 3345 (1993).

\bibitem{vN} J. von Neumann, {\em Mathematical Foundations of Quantum Mechanics} (Princeton University Press, Princeton, 1955).


\bibitem{hartlelo} J.B. Hartle,
{\em Spacetime Quantum Mechanics and the
Quantum Mechanics of Spacetime}
in `Gravitation and Quantizations',  in the
Proceedings of the 1992 Les Houches Summer School, ed. by B. Julia and J. Zinn-
Justin,  Les  Houches  Summer  School  Proceedings,
Vol. LVII, (North Holland, Amsterdam, 1995); [gr-qc/9304006].



\bibitem{ctp1}	J. S. Schwinger, {\em Brownian Motion of a Quantum Oscillator}, J. Math. Phys. 2, 407 (1961).

\bibitem{ctp2} L. V. Keldysh, {\em Diagram Technique for Nonequilibrium Processes}, Zh. Eksp. Teor. Fiz. 47, 1515 (1964).

\bibitem{ctp3} G. Zhou, Z. Su, B. Hao and L. Yu, {\em Equilibrium and nonequilibrium formalisms made unified}, Phys. Rep. 118, 1 (1985).

\bibitem{ctp4} B. S. DeWitt, {\em Effective action for expectation
values},  in "Quantum Concepts in Space and Time", ed. R. Penrose and
C. J. Isham (Claredon Press, Oxford, 1986)


\bibitem{ctp5}  E. Calzetta and B. L. Hu, {\em Nonequilibrium quantum fields: Closed-time-path effective action, Wigner function, and Boltzmann equation },  Phys. Rev.  D37, 2878 (1988).

\bibitem{ch87}  E. Calzetta and B. L. Hu, {\em Closed-time-path functional formalism in curved spacetime}, Phys. Rev.  D35, 495 (1987).

\bibitem{wein05} S. Weinberg,{\em Quantum contributions to cosmological correlations}
Phys. Rev. D 72, 043514 (2005).

\bibitem{CH08} E. A. Calzetta  and B. L. Hu, {\em Nonequilibrium quantum field theory} (Cambridge University Press, 2008).


\bibitem{Berges}  J. Berges,  {\em Introduction to nonequilibrium quantum field theory},  AIP Conf. Proc.  739,  3  (2004).

\bibitem{Berges2} J. Berges,  {\em Nonequilibrium Quantum Fields: From Cold Atoms to Cosmology}, Lecture Notes of the Les Houches Summer School: Strongly Interacting Quantum Systems out of Equilibrium (Oxford University Press, Oxford, 2016).

\bibitem{coma1}A. Kamenev, {\em Field Theory of Non-Equilibrium Systems} (Cambridge University Press, Cambridge, 2011).

\bibitem{coma2}  J. Rammer, {\em Quantum Field Theory of Non-equilibrium States} (Cambridge University Press, Cambridge, 2009)

\bibitem{Erice95} B. L. Hu,  {\em Correlation Dynamics of Quantum Fields and Black Hole Information Paradox},      talk at the International School of Astro-fundamental Physics, Sept. 	1995.  Proceedings edited by N. Sanchez and Zichichi 	(Kluwer Publishers, Dordrecht, 1996).

\bibitem{Jubb} I. Jubb, {\em Causal state updates in real scalar quantum field theory},
Phys. Rev. D 105, 025003 (2022).


\bibitem{nfsg}  E. Calzetta and B. L. Hu, `{\em Noise and Fluctuations in
Semiclassical Gravity'}, Phys. Rev.  D49, 6636 (1994).

\bibitem {HM3}
B. L. Hu and A. Matacz, {\em Backreaction in Semiclassical Gravity:
the Einstein-Langevin Equation}, Phys. Rev. 51 (1995).


\bibitem{LOCC} E. Chitambar, D. Leung, L. Mancinska, M. Ozols, and A. Winter, {\em Everything You Always Wanted to Know About LOCC (But Were Afraid to Ask)},  	Commun. Math. Phys., 328,  303 (2014).


\bibitem{Weinberg} S. Weinberg,  { \em The Quantum Theory of Fields: I. Foundations}  (Cambridge University Press, Cambridge,  1996).

\bibitem{QTPdet}
 C. Anastopoulos and N. Savvidou, {\em Measurements on relativistic quantum fields: I. Probability assignment}, arXiv:1509.01837.




\bibitem{hartlehu}  J. B. Hartle and B. L. Hu, {\em Quantum effects in the early universe. II. Effective action for scalar fields in homogeneous cosmologies with small anisotropy}, Phys. Rev.  D20, 1772
(1979).


\bibitem{FeVe1}  R. Feynman and F. Vernon, {\em The Theory of a general quantum system interacting with a linear dissipative system}, Ann. Phys. (NY) {\bf 24}, 118
(1963).

\bibitem{FeVe2} R. Feynman and A. Hibbs, {\em Quantum Mechanics and Path Integrals},
(McGraw - Hill, New York, 1965).

\bibitem{CaLe} A. O. Caldeira and A. J. Leggett, {\em Path integral approach to quantum Brownian motion},  Physica
{\bf 121A}, 587 (1983).



\bibitem{HPZ1}
B. L. Hu, J. P. Paz and Y. Zhang, {\em Quantum Brownian motion in a general environment: Exact master equation with nonlocal dissipation and colored noise}, Phys. Rev. {\bf D45}, 2843 (1992).




\bibitem{ML} J. C. Muga and J. R. Leavens, {\em Arrival time in quantum mechanics}, Phys. Rep. 338, 353 (2000).

 \bibitem{ToAbooks} J. C. Muga, R. S. Mayato, and I. L. Equisquiza, {\em Time in Quantum Mechanics, vol 1} (Springer 2008); J. G. Muga, A Ruschhaupt and A. Del Campo, {\em Time in Quantum
     Mechanics, vol 2} (Springer 2010).


\bibitem{Pauli}  W. Pauli, {\em The Principles of Quantum Mechanics}, in {\em Encyclopedia of Physics}, edited
by S. Flugge, Vol. 5/1 (Springer, Berlin, 1958).


\bibitem{Leon} J. Le\'on, {\em Time-of-arrival formalism for the relativistic particle }, J. Phys A: Math. Gen. 30, 4791 (1997).



   \bibitem{Kij} J. Kijowski,  {\em On the time operator in quantum mechanics and the Heisenberg uncertainty relation for energy and time}, Rep. Math. Phys. 6, 361 (1974).

\bibitem{Helfer}  A. D. Helfer,{\em The stress - energy operator},   Class. Quantum Grav. 13, L129 (1996).

\bibitem{Torre}   C. G. Torre and M. Varadarajan, {\em
Functional Evolution of Free Quantum Fields}, Class. Quantum. Grav. 16, 2651 (1999).

\bibitem{Kla85} JR Klauder and B Skagerstam, {\em Coherent States: Applications in Physics and Mathematical Physics} ( World Scientific, Singapore 1985).

\bibitem{berezin} F. A. Berezin, {\em The Method of Second Quantization} (Academic Press, 1966).



\bibitem{ByTa}  V.P. Bykov and V.I. Tatarskii, {\em Causality violation in the Glauber theory of photodetection}, Phys. Lett. A 136, 77 (1989).


\bibitem{Fleisch} M. Fleischhauer, {\em
Quantum-theory of photodetection without the rotating wave approximation} J. Phys. A: Math. Gen. 31, 453 (1998).



\bibitem{MJF95}P. W. Milonni, D. F. V. James, and H. Fearn, {\em Photodetection and causality in quantum optics}, Phys. Rev. A 52, 1525 (1995).



\bibitem{FoCo}  G. W. Ford and R. F. O'Connell, {\em The rotating wave approximation (RWA) of quantum optics: serious defect}, Physica A243, 377 (1997).






\bibitem{FCAH10} C. Fleming, N. I. Cummings, C. Anastopoulos and B. L. Hu,  {\em The Rotating-Wave Approximation: Consistency and Applicability from an Open Quantum System Analysis}, J. Phys. A: Math. Theor. 43, 405304 (2010).









\bibitem{QTP4}	C. Anastopoulos and N. Savvidou, {\em Quantum Temporal Probabilities in Tunneling Systems}, Ann. Phys. 336, 281 (2013).
\bibitem{QTP5}	C. Anastopoulos and N. Savvidou, {\em Path of a Tunneling Particle}, Phys Rev. A95 , 052120 (2017).


\bibitem{AnSav11}	C. Anastopoulos and N. Savvidou, {\em Coherences of Accelerated Detectors and the Local Character of the Unruh Effect}, J. Math. Phys. 53, 012107 (2012).


\bibitem{PapAna22} M. Papageorgiou and C. Anastopoulos, {\em Field observables recorded by Unruh-Dewitt detectors} (in preparation).


\bibitem{BaKa1} G. Baym and L. P. Kadanoff, {\em Conservation Laws and Correlation Functions}, Phys. Rev. 124, 287 (1961).

\bibitem{BaKa2}  G. Baym and L. P. Kadanoff, {\em Quantum Statistical Mechanics}, (Benjamin, New York 1962).




\bibitem{RH97} S. A. Ramsey, and B. L. Hu, B.L., {\em  O (N) quantum fields in curved spacetime}, Phys. Rev. D56, 661 (1997).



\bibitem{CH99} E. Calzetta  and B. L. Hu,  {\em Stochastic dynamics of correlations in quantum field theory: From the Schwinger-Dyson to Boltzmann-Langevin equation}, Phys. Rev. D61, 025012 (1999).


\bibitem {cddn} E. Calzetta and B. L. Hu, ``Correlations, Decoherence,
Disspation and Noise in Quantum Field Theory'', in {\it Heat Kernel
Techniques and Quantum Gravity}, ed. S. A. Fulling (Texas A\& M Press, College
Station 1995).

\bibitem{DCH}  E. Calzetta and B. L. Hu, ``Decoherence of Correlation
Histories'' in {\it Directions in General Relativity, Vol II: Brill
Festschrift}, eds B. L. Hu and T. A. Jacobson (Cambridge University Press,
Cambridge, 1993).



\bibitem{Ana97} C. Anastopoulos, {\em Coarse grainings and irreversibility in quantum field theory}, Phys. Rev. D56, 1009 (1997).



\bibitem{Ana06} C. Anastopoulos, {\em Classical versus quantum probability in sequential measurements}, Found. Phys. 36, 1601 (2006).

\bibitem{LeGa} A. J. Leggett and A. Garg, {\em Quantum mechanics versus macroscopic realism: Is the flux there when nobody looks?},
Phys. Rev. Lett. 54, 857 (1985).





\bibitem{Brunnt} J. Kofler and Č. Brukner, {\em Condition for macroscopic realism beyond the Leggett-Garg inequalities}, Phys. Rev. A 87, 052115 (2013).

\bibitem{Ana03} C.  Anastopoulos, {\em Quantum Correlation Functions and the Classical Limit}, C. Anastopoulos, Phys. Rev. D63, 125024 (2001).

\bibitem{Balescu}  R. Balescu,
{\em Equilibrium and Nonequilibrium Statistical Mechanics}
(John Wiley, New York, 1975).


\bibitem{MiSu77} B. Misra and E.C.G. Sudarshan,  {\em The Zeno’s Paradox in Quantum Theory}, J. Math. Phys. 18, 756 (1977).

\bibitem{HalYe} J. J. Halliwell and J. M. Yearsley, {\em Pitfalls of path integrals: Amplitudes for spacetime regions and the quantum Zeno effect}, Phys. Rev. D86, 024016 (2012).





\end{thebibliography}
\end{document}